\documentclass[11pt]{article}
\usepackage[super,compress]{cite}
\usepackage{graphicx}
\usepackage{caption,subcaption}
\usepackage{url,hyperref}
\usepackage{color}
\usepackage{amsmath,amsfonts,amssymb}

\def\locald{\rho_0}

\def\redN{\mu_n}
\newcommand{\sigsi}{\sigma^{SI}}
\newcommand{\sigsd}{\sigma^{SD}}

\newcommand{\mwimp}{m_\chi}
\def\msi{(\sigsi,\,\mwimp)}
\def\msd{(\sigsd,\,\mwimp)}
\def\sisd{(\sigsd,\,\sigsi)}

\begin{document}

\title{SCINTILLATING BOLOMETERS: A KEY FOR DETERMINING WIMP PARAMETERS}
\maketitle

\begin{center}
D. G. CERDE\~NO, C. MARCOS, M. PEIR\'{O}
\\Instituto de F\'{\i}sica Te\'{o}rica
      UAM/CSIC, Universidad Aut\'{o}noma de Madrid,  28049
      Madrid, Spain
\\Departamento de F\'{\i}sica Te\'{o}rica,
      Universidad Aut\'{o}noma de Madrid, 28049
      Madrid, Spain
\\
\vspace{0.5cm}
M. FORNASA
\\School of Physics and Astronomy, 
  University of Nottingham, 
  University Park, NG7 2RD, Nottingham, United Kingdom
\\
\vspace{0.5cm}
C. CUESTA\footnote{Present address: CENPA, University of Washington, US}, E. GARC\'IA, C. GINESTRA, M. MART\'INEZ\footnote{Fundaci\'on ARAID, C/ María de Luna 11, Edificio CEEI Arag\'on, 50018 Zaragoza, Spain. mariam@unizar.es}, Y. ORTIGOZA, J. PUIMED\'ON, M. L. SARSA
\\Grupo de F\'\i sica Nuclear y Astropart\'\i culas,
      Universidad de Zaragoza, 
      50009 Zaragoza, Spain
\\Laboratorio Subterr\'aneo de Canfranc, Paseo de los Ayerbe s.n., 22880 Canfranc Estaci\'on, Huesca, Spain
\end{center}

\begin{abstract}
In the last decade direct detection Dark Matter (DM) experiments have 
increased enormously their sensitivity and ton-scale setups have been proposed, 
especially using germanium and xenon targets with double readout and 
background discrimination capabilities. 
In light of this situation, we study the prospects for determining the parameters of
Weakly Interacting Massive Particle (WIMP) DM (mass, spin-dependent (SD) and 
spin-independent (SI) cross section off nucleons) by combining the results of 
such experiments in the case of a 
hypothetical detection.
In general, the degeneracy between the SD and SI components of the scattering cross section can only be removed using targets with different sensitivities to these components.
Scintillating bolometers, with particle discrimination 
capability, very good energy resolution and threshold and a wide choice of 
target materials, are an excellent tool for a multitarget complementary DM search.
We investigate how the simultaneous use of scintillating targets with 
different SD-SI  sensitivities and/or light isotopes (as the case of 
CaF$_2$ and NaI) significantly improves the determination of the WIMP
parameters.
In order to make the analysis more realistic we include  
the effect of uncertainties in the halo model and in 
the spin-dependent nuclear structure functions, as well as the effect of a 
thermal quenching different from 1.
\end{abstract}


\section{Introduction}	
\label{sec:intro}
Weakly Interacting Massive Particles (WIMPs) can be directly detected through 
their scattering off target nuclei of a detector\cite{goodman85}. 
In the last decades, numerous experiments, using different targets and 
detection techniques, have been searching for WIMPs or are currently taking 
data. Some of them have searched for distinctive signals, such as an annual 
modulation in the detection rate: DAMA\cite{bernabei2003} and DAMA/LIBRA\cite{bernabei2008, bernabei2013}, 
using NaI scintillators, have reported a highly significant signal
(9.3$\sigma$) and CoGeNT\cite{Aalseth:2014eft, Aalseth:2014jpa} claimed a less significant 
evidence (2.2$\sigma$) in the first three years of its data, gathered with a Ge
semiconductor. 
Moreover, CoGeNT\cite{Aalseth2011}, CRESST\cite{Angloher2012} (using CaWO$_4$ 
scintillating bolometers) and CDMS II (with data from its Si detectors)\cite{PhysRevLett.111.251301} have reported  
excesses of events at low energies that could be compatible with a signal produced by light 
WIMPs with a mass of the order of 10~GeV. On the other hand, 
XENON10\cite{Angle2011}, XENON100\cite{Aprile2012}, 
LUX\cite{2013arXiv1310.8214L} (also based on Xe), the abovementioned 
CDMS II \cite{Ahmed2010}, EDELWEISS\cite{Armengaud2011, Ahmed2011} (with Ge), 
KIMS\cite{Kim2012} (with CsI), 
PICASSO\cite{Archambault2012} (with C$_4$F$_{10}$), SIMPLE\cite{Felizardo2012} 
(with C$_2$ClF$_5$) and COUPP\cite{Behnke2011} (with CF$_3$I) have obtained 
negative results setting more stringent upper bounds on the 
WIMP-nucleon cross sections.
Currently the strongest limits are obtained by the LUX 
collaboration, excluding spin-independent WIMP-nucleon elastic scattering 
cross sections larger than 7.6$\times$10$^{-46}$~cm$^2$ for a WIMP mass of 33 
GeV, and the SuperCDMS collaboration for low mass WIMPs \cite{Agnese:2013jaa, Agnese:2014aze}.
In the next years new experiments and upgraded versions of the existing ones 
are going to explore even smaller cross sections, closing in on DM searches.

The final goal of all these experiments is to determine the nature of DM,
measuring some of its properties (namely its mass and interaction cross section with 
ordinary matter). Signals from different targets are needed, since they can provide complementary information which can lead to a better determination of the DM parameters.\cite{Bertone2007,Pato2011} 
In a previous paper\cite{cerdeno2013a} we analysed the complementarity of a 
Ge and a Xe experiment with energy thresholds and resolutions already achieved by CDMS 
and XENON100 experiments, respectively, and with background levels expected for their 
corresponding extensions (SuperCDMS\cite{Sander:2012nia} and 
XENON1T\cite{Aprile:2012zx}). For different WIMP scenarios, we assumed
hypothetical detections with an exposure of 300~kg$\times$yr in both 
experiments and we concluded that the combination of data from Xe and 
Ge-based detectors might not lead to a good reconstruction of all the WIMP parameters,
since there is a degeneracy in the SI and SD parts of the scattering WIMP-nucleus cross section, and both targets have very similar SI over SD sensitivity (see also Ref.\,\citen{Newstead:2013pea} for a recent study on the non-complementarity of Xe and Ar). We showed that incorporating targets with different sensitivities to SI and SD interactions 
could significantly improve the 
reconstruction. We considered the case of some of the most promising 
scintillating bolometric targets: CaWO$_4$ (currently used by CRESST), Al$_2$O$_3$ and
LiF (studied by ROSEBUD\cite{2010idm..confE..54C}, that could be considered 
in the future as additional targets in EURECA\cite{Kraus:2011zz}, a European 
collaboration that plans to search for WIMPs with a 1-ton cryogenic hybrid 
detector). 

We observed that the inclusion of CaWO$_4$ (being mainly sensitive
to SI couplings) only leads to a total complementary result for a WIMP of 50~GeV in a small 
region of the plane ($\sigsi,\sigsd$)  
in which the expected events in Ge and Xe are 
mainly due to SD interactions. 
On the other hand, Al$_2$O$_3$ and LiF (being more sensitive to SD 
interactions) achieve complementarity with germanium and xenon in regions of 
the parameter space where the rate in the latter is dominated by SI 
couplings. We also determined the exposures and background levels required 
by the bolometers to be complementary to Ge- and Xe-based experiments.

In this paper we follow the same strategy and reanalyze the role of Ge- and 
Xe-based experiments in light of the improved (or potential) energy thresholds
in CDMS and LUX
\footnote{Notice that a threshold as low as 2~keV has been reported in previous CDMS II analysis \cite{Ahmed:2010wy} although not for a background free search. In order to simplify the comparison with LUX, we will here assume the same threshold of 3~keV, considering that the new iZIP detectors in SuperCDMS might allow a much better background subtraction.}.
We also study the complementarity with two additional targets: CaF$_2$ and 
NaI. The first one has already been used as scintillating bolometer\cite{Alessandrello1992,Bobin1997},
 whereas the construction of a bolometer based on
NaI (which is a hygroscopic and fragile material) is an ongoing R\&D 
project of the Zaragoza group.\cite{Coron2013}
We include in our analysis not only the effect of the previously considered uncertainties in the halo 
parameters and SD structure functions, but also the possible influence
of the thermal quenching between nuclear and electron recoils in the complementarity of these targets.

The structure of this article is as follows: Sec.~\ref{sec:wimpParam} 
is a short summary of the methodology we follow in reconstructing the 
WIMP parameters from the (simulated) data in direct detection 
experiments. In Sec.~\ref{sec:uncertainties} we address the most relevant 
uncertainties in the analysis, in particular the astrophysical ones 
(due to our imperfect knowledge of the DM halo of the Milky Way), 
those related to the SD Structure Functions (SDSF) parametrizing the spin content
of the nucleons in the target and, finally, the effect of changing the 
thermal quenching $q$.
In Sec.~\ref{sec:geXe} we present the results for some selected benchmarks
when considering only Ge and Xe experiments, finding that the combination of data from these two targets contributes to a better measurement of the WIMP parameters, but a degeneracy in the SD and SI independent cross section usually remains. In 
Sec.~\ref{sec:scintBolo} we describe the characteristics of the scintillating 
targets under study (i.e. CaF$_2$ and NaI ). In
Sec.~\ref{sec:results} we show how their inclusion can lead to a better 
determination of the DM mass and scattering cross section, breaking in some cases the SI-SD degeneracy.
Finally, conclusions are presented in Sec.~\ref{sec:conclusions}.

\section{Reconstructing WIMP parameters from signals in direct detection experiments}
\label{sec:wimpParam}
In the standard analysis framework for WIMP direct 
detection\cite{Smith1990,Lewin1996} (see also Refs.\,\citen{Cerdeno:2010jj} 
and \citen{Peter:2013aha} for recent reviews) the WIMP-nucleus scattering 
cross section is separated into a SI and a SD contribution, with $f_p$ and
$a_p$ being the corresponding effective couplings to protons and $f_n$ and
$a_n$ to neutrons. In order to reduce the number of parameters that 
characterize the expected event rate, we assume here that the SI coupling 
is isospin-invariant ($f_p = f_n$) and we take a specific relation between 
$a_p$ and $a_n$ (namely $a_p / a_n =-1$). 
Under these assumptions, the generic WIMP is completely determined by its mass 
$\mwimp$, the SI contribution to the WIMP-nucleon cross section $\sigsi$ and 
the SD component $\sigsd$. 

Thus, the total number of WIMP recoil events in a given energy window can be 
expressed as
\begin{equation}
    N=\sum_{isotopes}f({\cal C}_{ SI}~{\sigsi}+{\cal C}_{ SD}4(S_p-S_n)^2{\sigsd})\,,
\label{eq:N}
\end{equation}
where, for each isotope, $f$ is its mass fraction in the 
detector, $S_p$ and $S_n$ are the expectation values of the total spin 
operators for protons and neutrons respectively and the coefficients 
${\cal C}_{ SI}$ and ${\cal C}_{ SD}$ can be written as follows,
\begin{equation}
\begin{split}
    {\cal C}_{ SI}\equiv \int dE_R\int \left( \frac{\epsilon\locald f(v)}{2\redN^2 \mwimp v} \right)\,A^2\,F^2_{ SI}\,dv\,, \\
    {\cal C}_{ SD}\equiv \int dE_R\int \left( \frac{\epsilon\locald f(v)}{2\redN^2 \mwimp v} \right)\, \left( \frac{J+1}{3J} \right) \,F^2_{ SD}\,dv\,.
\end{split}
\label{eq:Cs}
\end{equation}
$\epsilon$ is the experimental exposure, $\locald$ is the local WIMP density, 
$f(v)$ is the WIMP speed distribution in the Earth reference frame normalized to unity, $\redN$ is the 
WIMP-nucleon reduced mass, $E_R$ is the nucleus recoil energy, $F^2_{ SI}$ 
($F^2_{ SD}$) is the SI (SD) nuclear form factor, $A$ is the nucleus mass 
number, and $J$ its nuclear spin.

We focus the analysis on two benchmark cases. For each of 
them (and for each target independently) we calculate the signal that such WIMPs
would produce in that specific detector by computing the number of recoil 
events, $\{\lambda_i\}^a$, expected for target $a$ in the $i$-th bin of $N$ evenly-spaced energy bins contained
in the energy window for WIMP search of each experiment. These expected 
events, $\{\lambda_i\}^a$, represent our experimental data, $\mathbf{D}$, and we are 
interested in estimating how well such simulated measurements can
be used to reconstruct the WIMP parameters.

In order to do so, we perform scans over the parameter space 
($\mwimp=1-10^5$~GeV, $\sigsi=10^{-12}-10^{-6}$~pb, and $\sigsd=10^{-8}-1$~pb).
For every point in the scan we compute the number of recoil events $N_i^a$ in the $i$-th energy bin for every target $a$ and then 
compute the likelihood comparing $N_i^a$ with the prediction of the benchmark model in the same energy bin for the same target, 
assuming that data from each experiment follow independent Poissonian distributions. 
We present the results as 68\% and 99\% confidence regions in the profile likelihood (PL). 
The nuclear and astrophysical uncertainties are considered as nuisance parameters. 
The scans are performed with MultiNest 3.0\cite{Feroz2009} interfaced with 
our own code for the computation of the number of recoil events and of the 
likelihood. Logarithmic flat priors are assumed for the three variables. 
We refer to Ref.\,\citen{cerdeno2013a} for a detailed description of how
the scans are performed.

If only one target is considered, the reconstruction of WIMP parameters is 
affected by degeneracies, since the number of events detected can be 
explained by different combinations of ($\mwimp$, $\sigsi$, $\sigsd$).
Such degeneracy can be broken by including more targets in the analysis: 
in Ref.\,\citen{cerdeno2013a} we defined ``complementarity'' as the situation in 
which a certain set of experiments manages to determine $\mwimp$, $\sigsi$ and 
$\sigsd$ with a certain finite accuracy, or, equivalently, when 68\% 
confidence level of the 2-dimensional contours are closed simultaneously in the three
planes ($\mwimp$, $\sigsi$), ($\mwimp$, $\sigsd$) and ($\sigsi,\sigsd$).

The following two WIMP benchmarks will be considered in the remaining
sections: 
\begin{itemize}
\item VL-SI: $\mwimp$=20~GeV, $\sigsi$=10$^{-9}$~pb, $\sigsd$=10$^{-5}$~pb, 
corresponding to a very light WIMP for which the SI contribution dominates the detection rate in Ge and Xe,
\item L-SD: $\mwimp$=50~GeV, $\sigsi$=10$^{-10}$~pb, 
$\sigsd$=1.5$\times$10$^{-4}$~pb, a light WIMP for which the SD contribution dominates in Ge and Xe.
\end{itemize}


\subsection{Including uncertainties}
\label{sec:uncertainties}

The expected DM signal depends on parameters affected by large uncertainties. In the following, we will take into account uncertainties in the velocity distribution of DM in the Milky Way halo, the spin dependent form functions for the target nuclei and the performance of the detector.

We considered a velocity distribution function that 
differs from the standard halo model by the presence of a high-velocity tail. 
Such a model, adopted from Ref.\,\citen{Lisanti:2010qx}, is well motivated by
N-body simulations and the velocity distribution can be written as follows,
\begin{equation}
F(v)=N_k^{-1}v^2[e^{-v^2/kv_0^2}-e^{-v_{esc}^2/kv_0^2}]^k\Theta(v_{esc}-v)\,,
\label{eq:halo}
\end{equation}
where $N_k=v_0^3e^{-y_e^2}\int_0^{y_e}y^2(e^{-(y^2-y_e^2)/k}-1)^kdy$, $y_e=v_{esc}/v_ 0$
and $k$ is the parameter that quantifies the deviation from the standard halo 
model, recovering it in the limiting case of $k$=0\cite{Lisanti:2010qx}.
This expression for the velocity distribution depends on three parameters: 
$v_{esc}$, $v_0$ and $k$. In order to account for our 
ignorance on the true velocity distribution of the DM in the halo of our Galaxy
we leave such parameters free to vary within the following ranges: $v_{esc}= [478,\, 610]$~km/s, $v_0=[170,\, 290]$~km/s, and $k=[0.5,\, 3.5]$. We also scan over
the local WIMP density $\locald$, in the range between 0.2 and 0.6 GeV/cm$^3$. All these parameters are subject to a uniformly flat prior distribution.

Regarding the WIMP interaction with the nucleus, it has been shown in 
Ref.\,\citen{cerdeno2013} that uncertainties in the SI and SD form factors of 
the target nuclei play a very different role. In the case of SI interactions, 
differences in the form factor can be safely neglected (in the present paper 
we have used the Helm form factor). On the other hand, for SD 
interactions the expressions of the form factors are more dependent on the nuclear model. 
These differences can significantly affect the expected
WIMP rate and, thus, the reconstruction of the WIMP parameters (especially
when SD interactions play a relevant role). 

\begin{figure}[t!]
 \hspace*{-0.5cm}
\begin{subfigure}[b]{0.37\textwidth} \includegraphics[width=\textwidth]{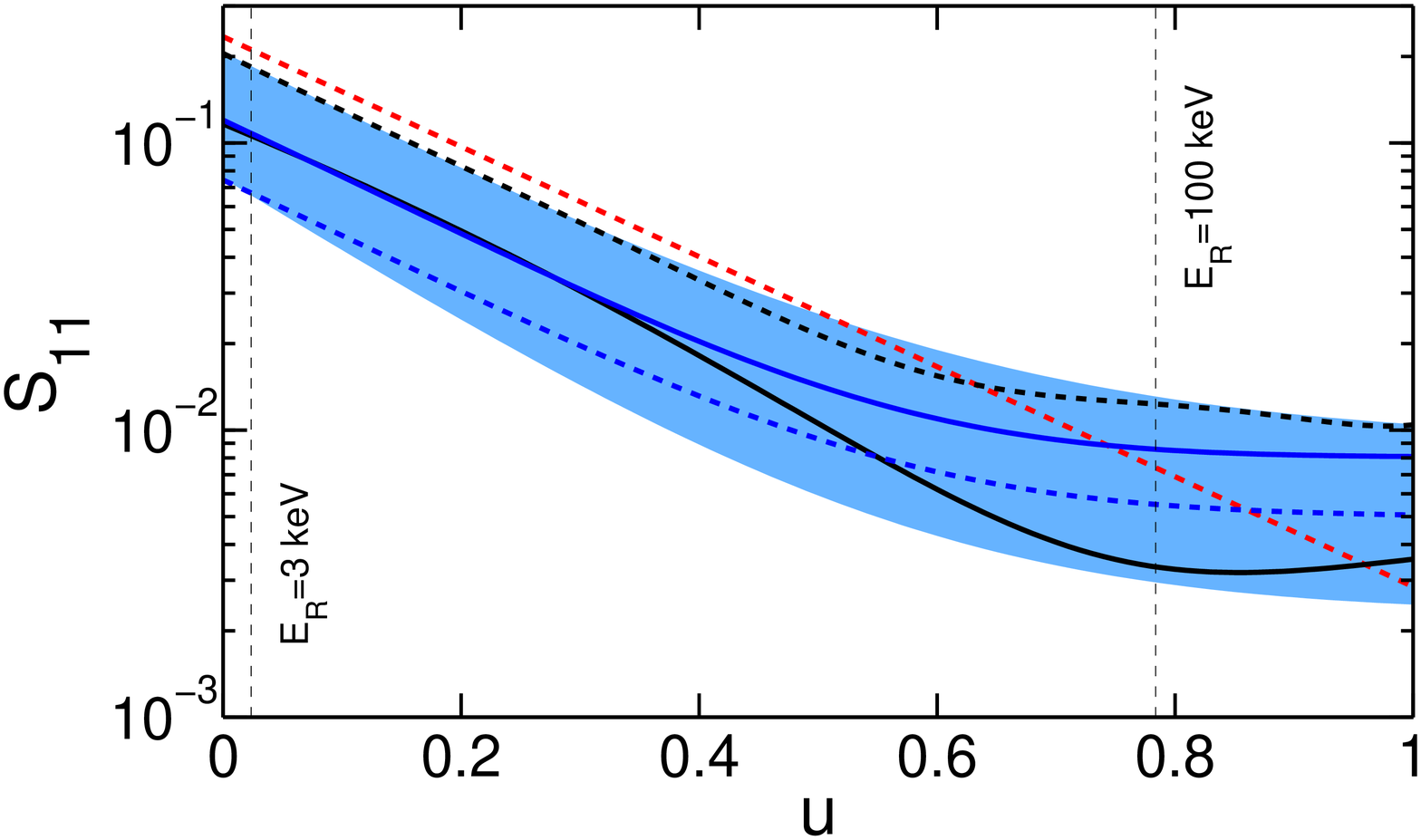} \caption{$^{73}$Ge}  \end{subfigure}\hspace*{-0.45cm}
\begin{subfigure}[b]{0.37\textwidth} \includegraphics[width=\textwidth]{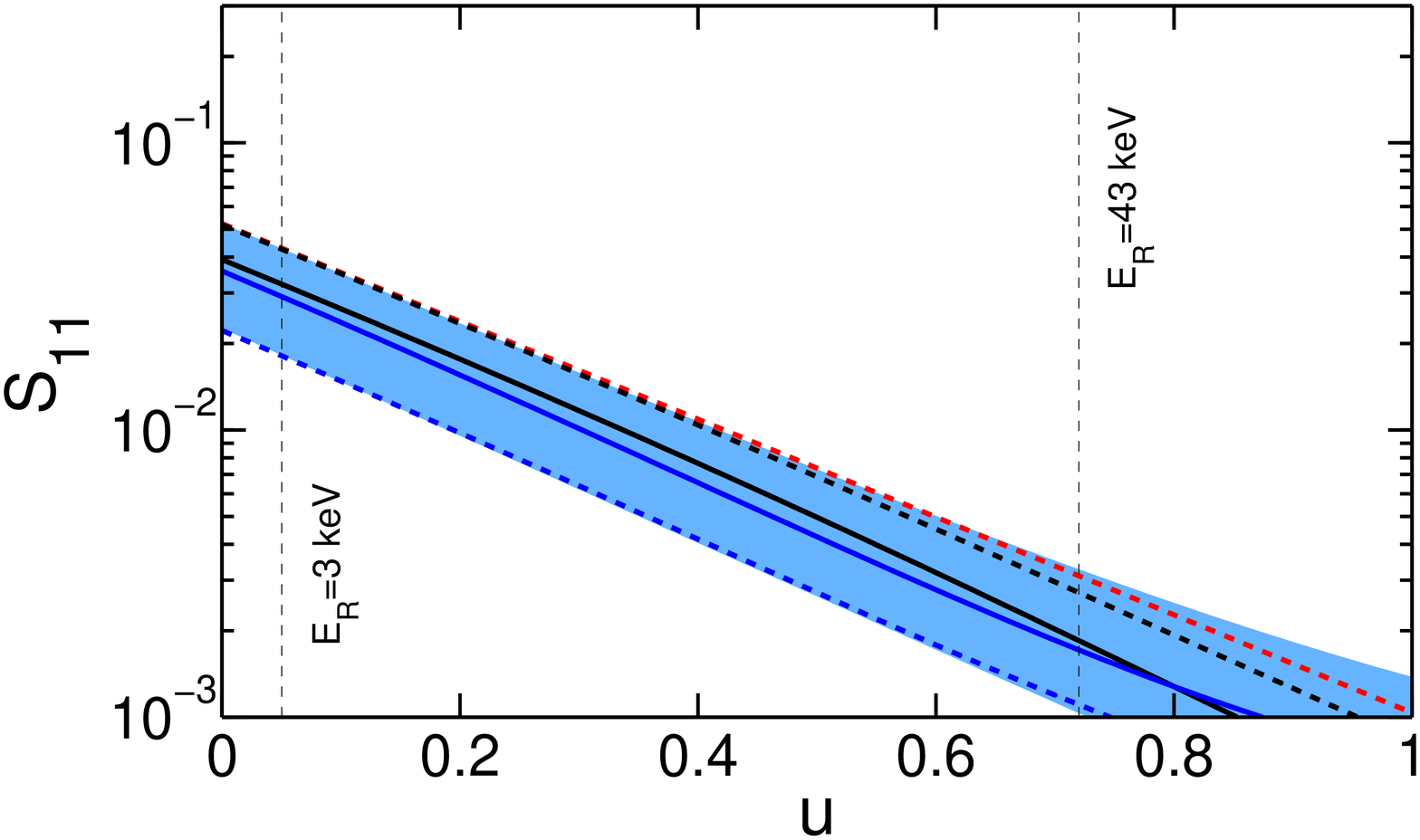} \caption{$^{129}$Xe}  \end{subfigure}\hspace*{-0.45cm}
\begin{subfigure}[b]{0.37\textwidth} \includegraphics[width=\textwidth]{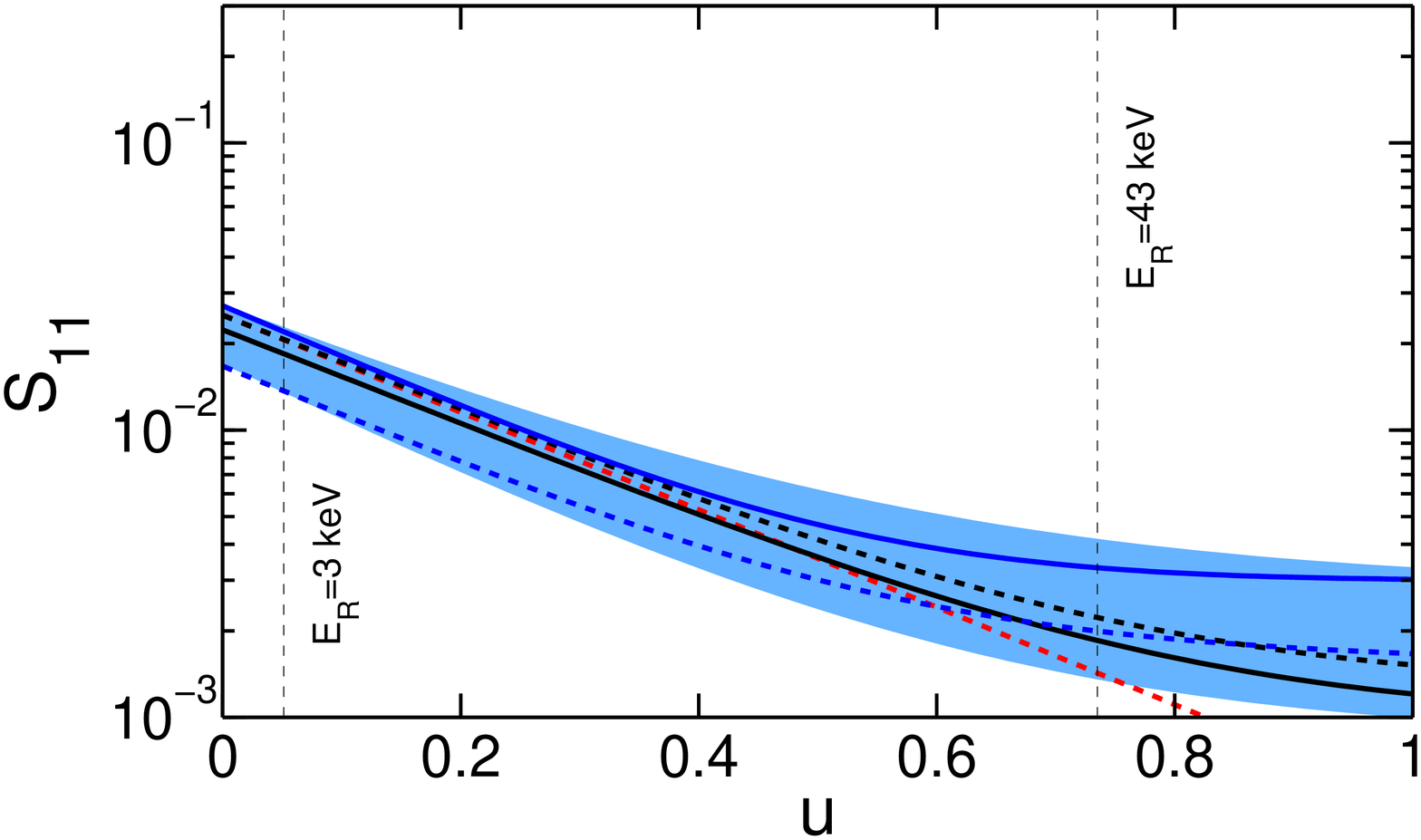} \caption{$^{131}$Xe}  \end{subfigure}\hspace*{-0.45cm}
  \\[0.0cm]
 \hspace*{-0.5cm}
\begin{subfigure}[b]{0.37\textwidth} \includegraphics[width=\textwidth]{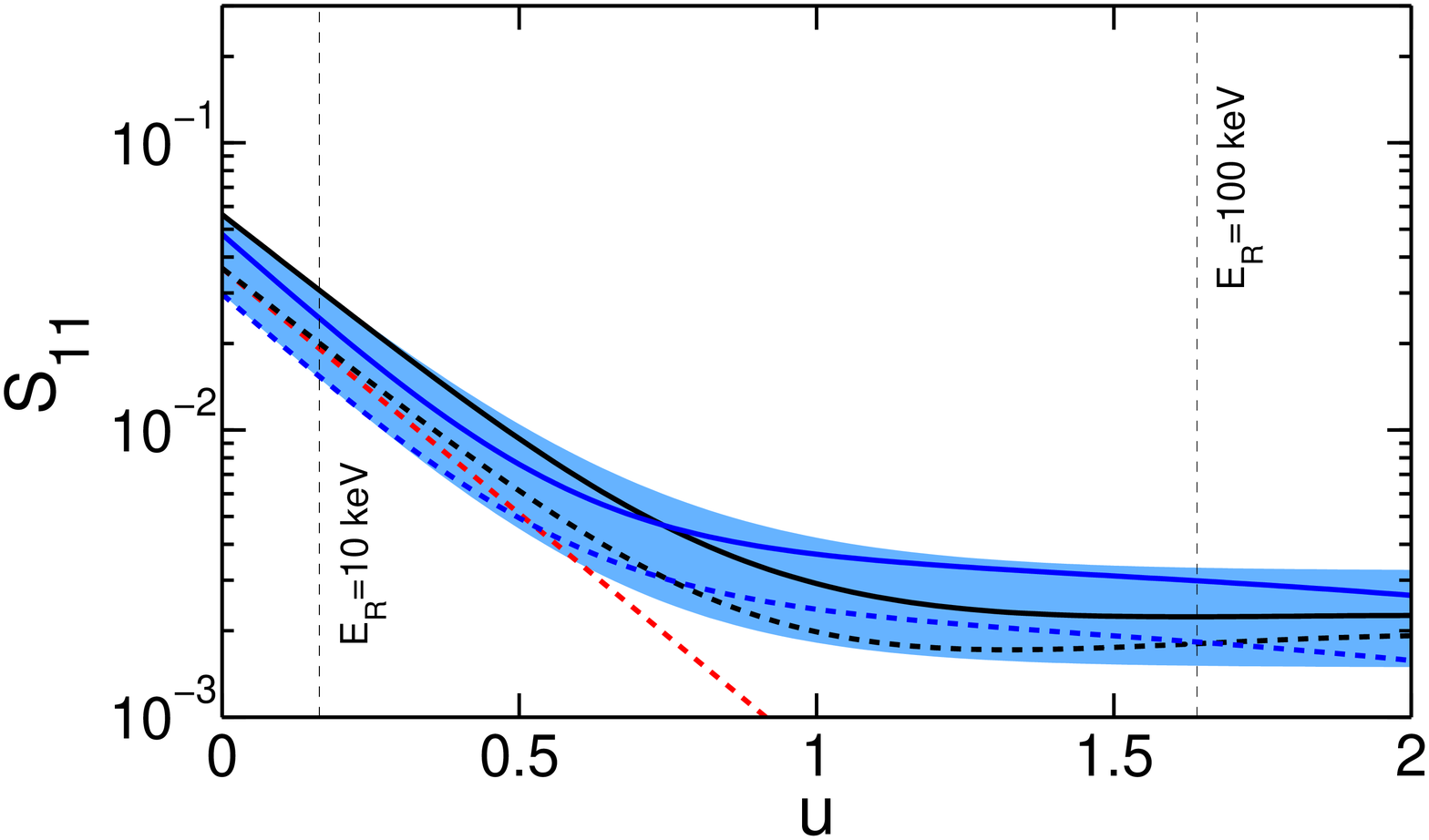} \caption{$^{127}$I}  \end{subfigure}\hspace*{-0.45cm}
\begin{subfigure}[b]{0.37\textwidth} \includegraphics[width=\textwidth]{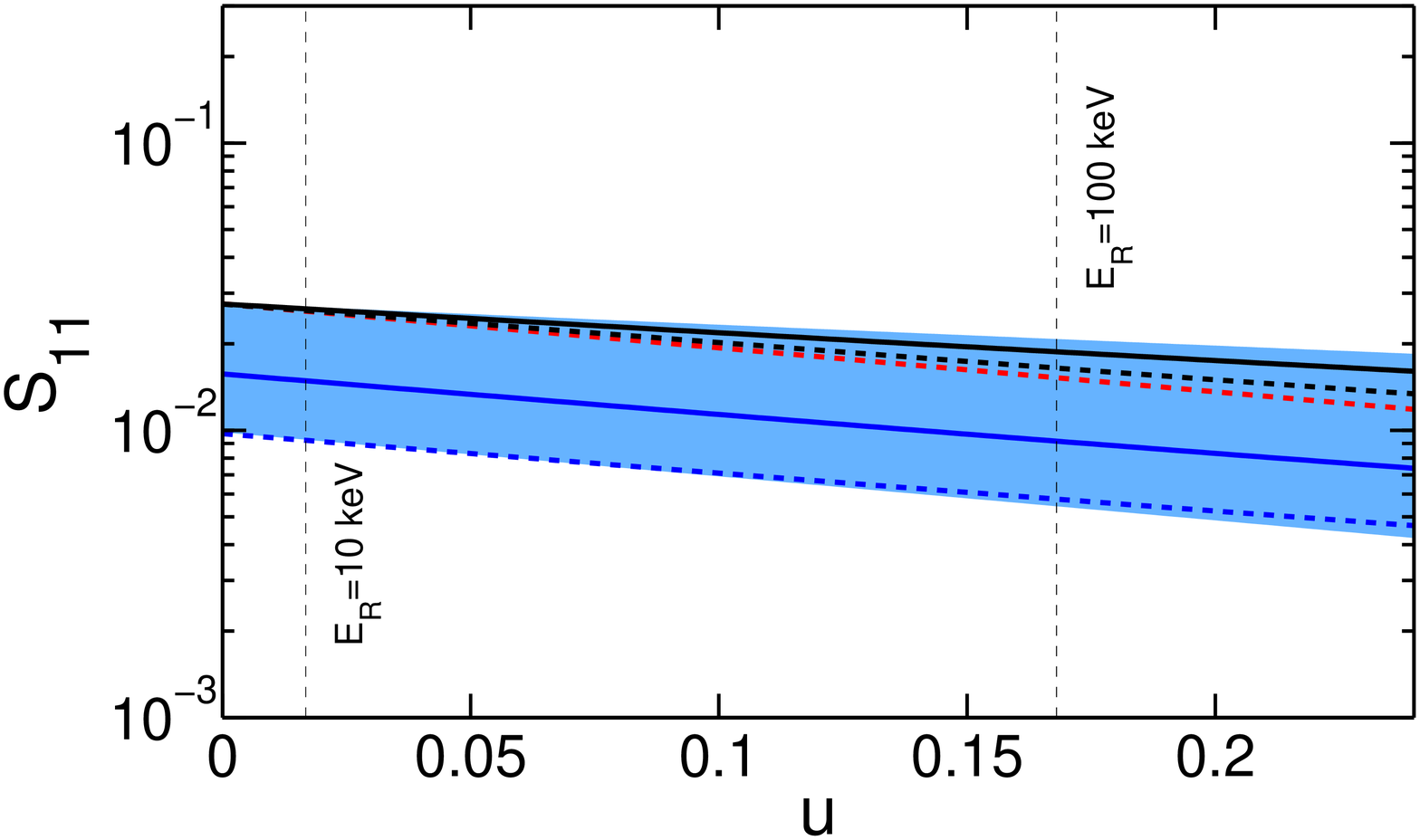} \caption{$^{23}$Na}  \end{subfigure}\hspace*{-0.45cm}
\begin{subfigure}[b]{0.37\textwidth} \includegraphics[width=\textwidth]{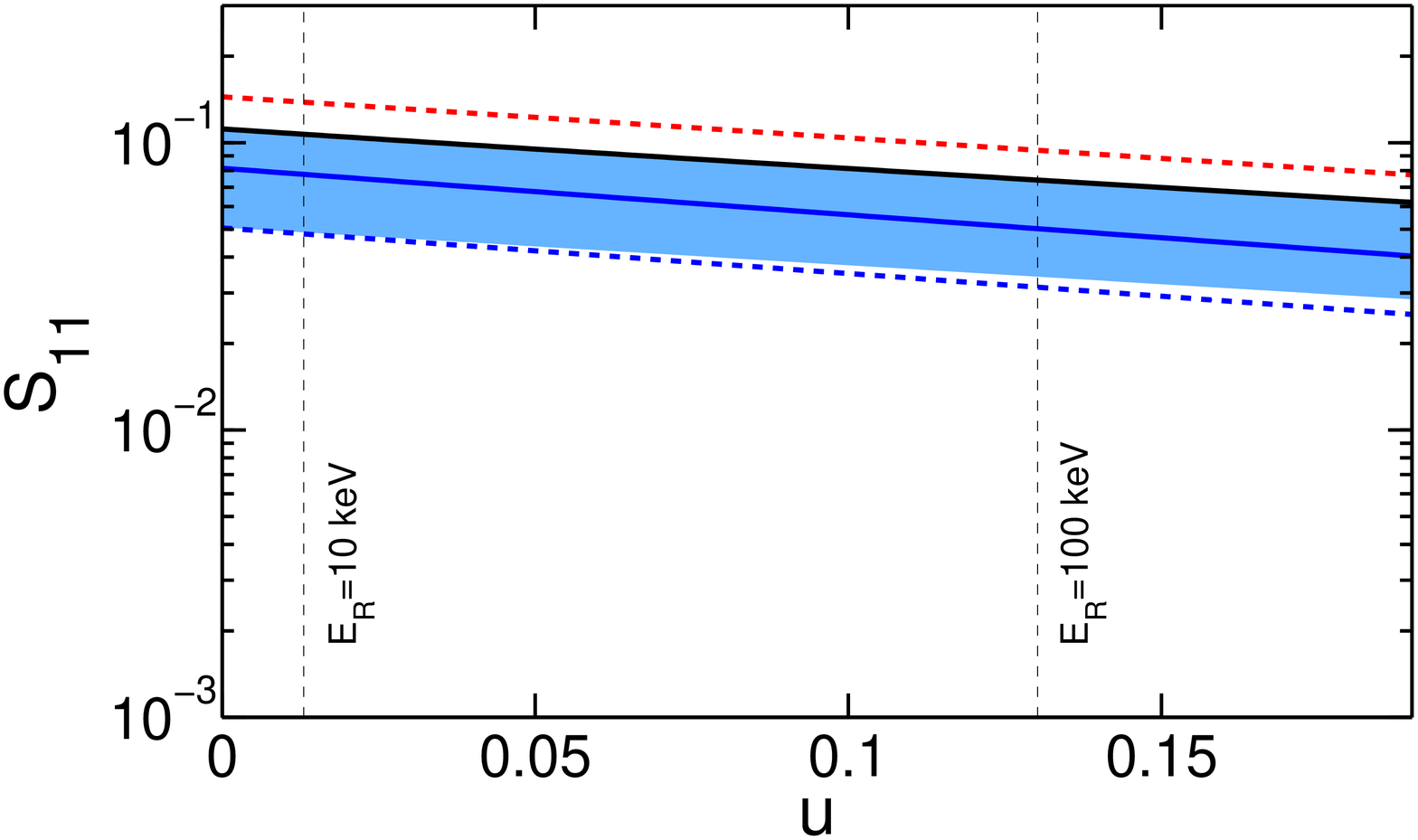} \caption{$^{19}$F}  \end{subfigure}\hspace*{-0.45cm}
\caption{Light blue area: SDSFs as a function of $u=(qA^{1/6})^2/2$ for $a_p$/$a_n=-1$ from Eq.~\ref{eq:sdUnc} (parameters varying within the ranges of Tab.~\ref{tab:SfSD}). Blue dashed (solid) line: Klos et al. min (max) model\cite{Klos:2013rwa}. Red dashed line: gaussian approximation\cite{Belanger:2008sj}. a) Black dashed (solid) line: Resell et al.\cite{Ressell:1993qm} (Dimitrov et al.\cite{Dimitrov:1994gc}). b,c,d) Black dashed (solid) line: Bonn (Nijm)\cite{Ressell:1997kx}. e) Black dashed (solid) line: Resell-Dean\cite{Ressell:1997kx} (Vergados et al.\cite{Divari:2000dc}). f) Black solid line: Vergados et al.\cite{Divari:2000dc}
}
\label{fig:SfSD}
\end{figure}

To take into account such uncertainties, the SDSFs are parametrized as follows,\cite{cerdeno2013}
\begin{equation}
  S_{ij}(u)=N((1-\beta)e^{-\alpha u}+\beta),
  \label{eq:sdUnc}
\end{equation}
where $u$ is an adimensional quantity proportional to the square of the 
momentum transfer, $u=(qb)^2/2$, in terms of the oscillator size parameter $b=A^{1/6}$. 
Note that for the case $a_p$/$a_n=-1$ the only contribution comes from the $S_{11}$ SDSF.
Tab.~\ref{tab:SfSD}  shows the ranges in which the 
three parameters $N$, $\alpha$ and $\beta$ have been allowed to vary for
each nucleus in order to reproduce the various determinations of the form 
factors available in the literature (see Ref.\,\citen{cerdeno2013} for a 
detailed explanation). Results for the SDSFs of the isotopes relevant in 
this work in the case $a_p$/$a_n=-1$ are displayed in Fig.~\ref{fig:SfSD} 
(light blue area) together with the most relevant nuclear calculations.
\\
\begin{table}
\begin{center}
\begin{tabular}{|c|c|c|c|}
\hline
Isotope & N & $\alpha$ & $\beta$ \\ 
\hline
$^{73}$Ge & 0.0749 - 0.2071 & 5.0 - 6.0 & 0.0304 - 0.0442\\ \hline
$^{129}$Xe & 0.0225 - 0.0524 & 4.0625 - 4.3159 & 0.001 - 0.0093\\ \hline
$^{131}$Xe & 0.0169 - 0.0274 & 3.9913 - 4.7075 & 0.05 - 0.105\\ \hline
$^{127}$I & 0.0297 - 0.0568 & 4.0050 - 4.4674 & 0.05 - 0.057\\ \hline
$^{23}$Na & 0.0098 - 0.0277 & 2.0 - 3.5287 & 0 - 0.1250\\ \hline
$^{19}$F & 0.0505 - 0.1103 & 2.9679 - 3.0302 &  0 - 0.0094 \\ \hline
\end{tabular}
\caption{Ranges considered for the parameters $N$, $\alpha$ and $\beta$ that describe the SDSFs (see Eq.~\ref{eq:sdUnc}) for the isotopes studied in this work.}
\label{tab:SfSD}
\end{center}
\end{table}

Finally, important systematics can also arise from 
the detection technique itself. Among these, we consider the effect of the thermal 
quenching factor, $q$, that measures the relative efficiency 
in the conversion into measurable thermal signal of the nuclear 
recoils energy deposition with respect to that corresponding to electron recoils, 
since the detectors are calibrated with gamma 
sources and the measured spectra are given in electron-equivalent energy.
This factor is typically assumed equal to one for bolometers but small 
deviations (of about 10-15\%) have been measured in different detectors (see 
for example Ref.\,\citen{Coron2008a} and references therein).
To illustrate the influence of this uncertainty on $q$, we consider here 
three different values ($q$ = 0.85, 1, and 1.15) for the NaI target.
\subsection{Results for Ge and Xe}
\label{sec:geXe}
As in our previous paper\cite{cerdeno2013a}, we start the analysis studying 
the complementarity of two experiments, based, respectively on Ge and Xe. Such
elements are employed by the collaborations producing the currently most 
stringent limits on WIMP properties and are contemplated in projects planning 
to extend the search to the ton scale (e.g. EURECA, 
SuperCDMS and XENON1T) or even to the multi-ton scale 
(LZ\cite{Malling:2011va} and DARWIN\cite{Baudis:2012bc}). 
Consequently, these targets are expected to represent the most sensitive 
experiments (at least in the most general WIMP scenarios) in the near future.

For our study, we have assumed a positive result (WIMP detection) in two 
experiments, one using a Ge-based target and the other using Xe. We consider 
the two detections combined when reconstructing the WIMP parameters. 
The same exposure ($\epsilon$ = 300 kg$\times$yr) is assumed for both 
experiments, as well as zero background\footnote{In our previous work \cite{cerdeno2013a} we checked that the expected backgrounds for SuperCDMS and XENON1T are so low that have no impact in the results, so zero background can be safely assumed.}. 
The energy window is set to [3-100] 
keV for Ge and [3-43] keV for Xe, where the lower values account for the 
recent or potential improvements in nuclear recoil energy thresholds of some Xe and Ge experiments. 
Tab.~\ref{tab:bm} shows the expected number of WIMP recoil events for the 
considered benchmarks over the whole energy range.

\begin{figure}[h]
 \hspace*{-0.5cm}
    \includegraphics[width=0.37\textwidth]{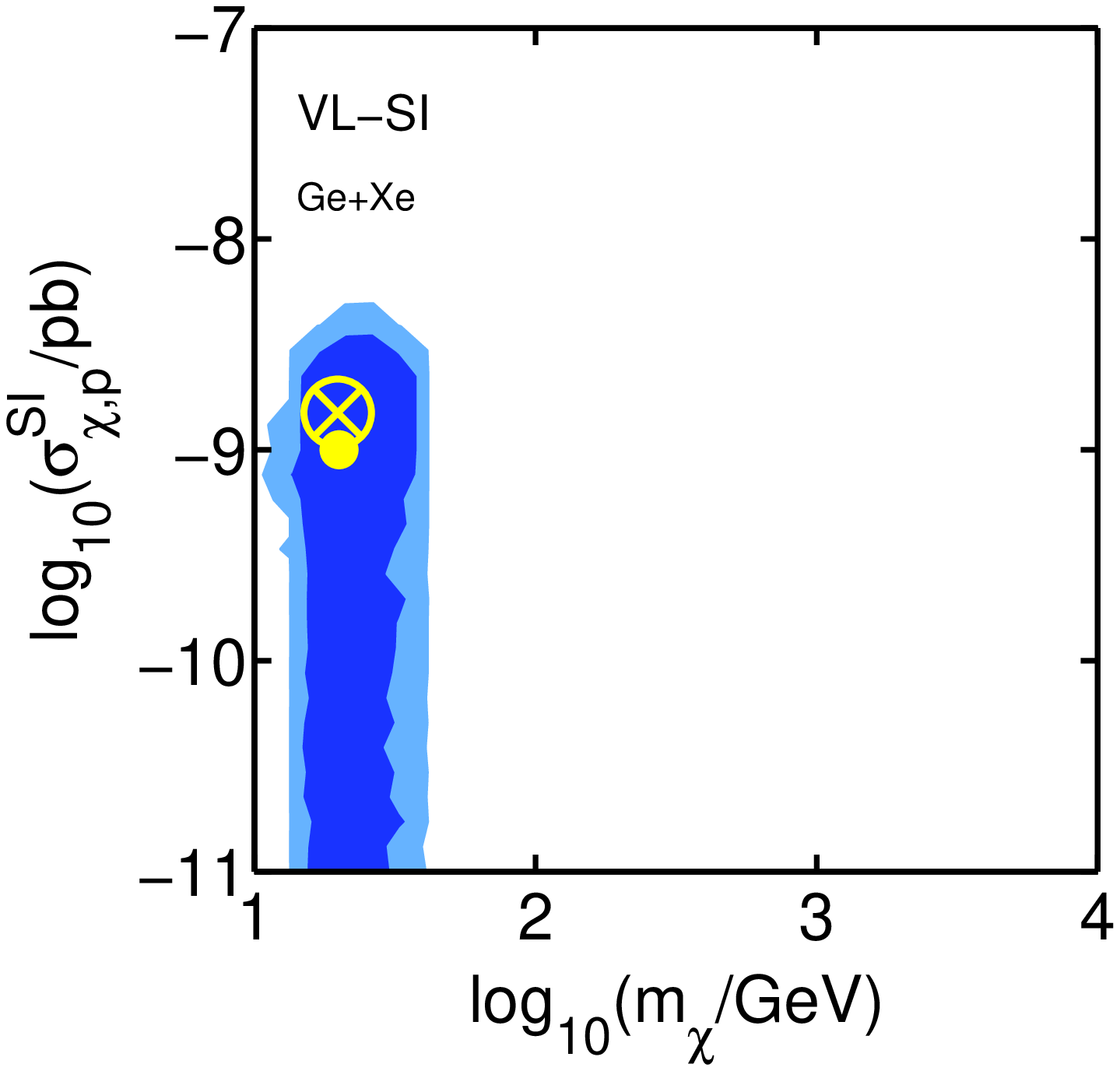}\hspace*{-0.45cm}
    \includegraphics[width=0.37\textwidth]{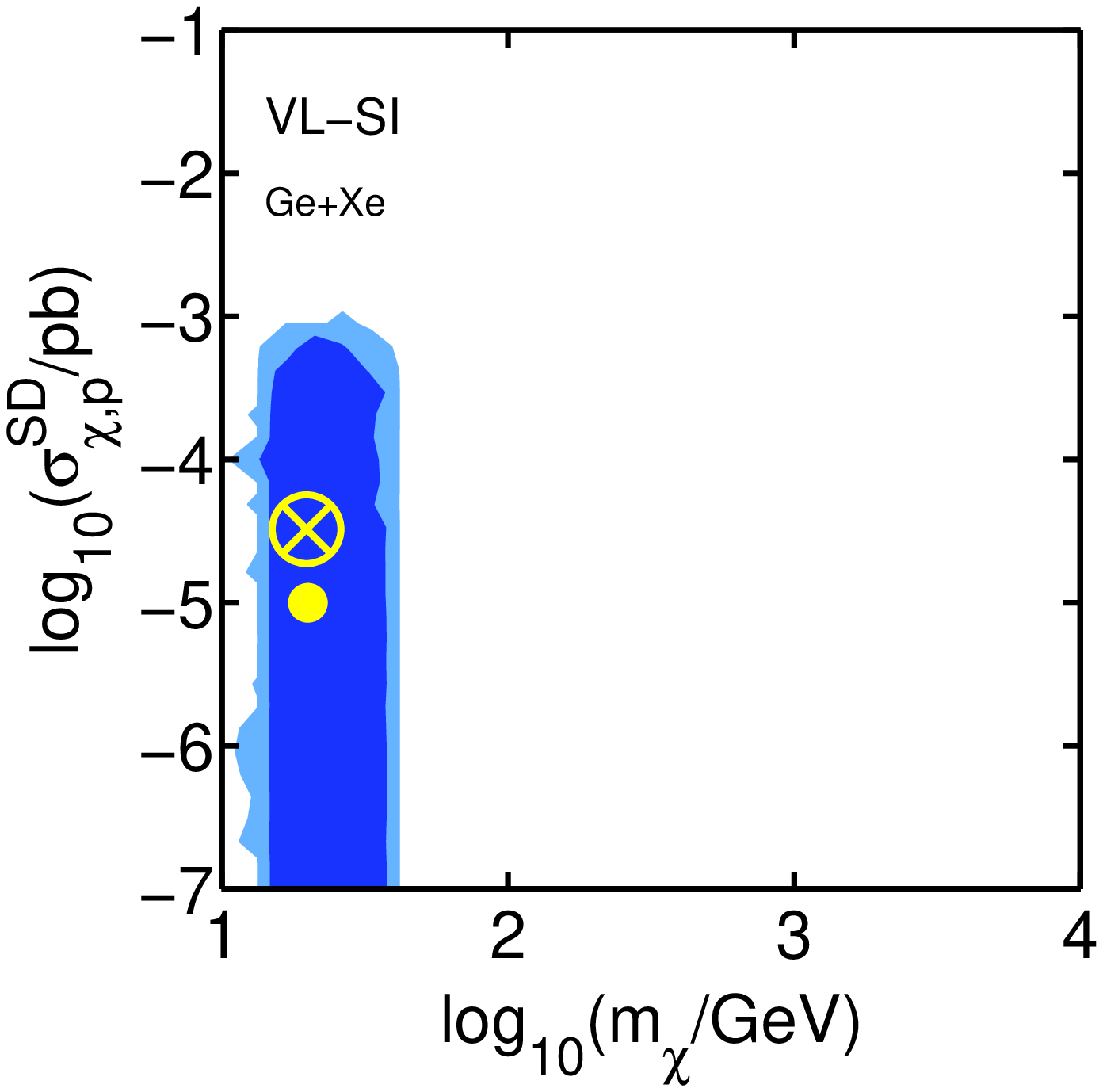}\hspace*{-0.45cm}
    \includegraphics[width=0.37\textwidth]{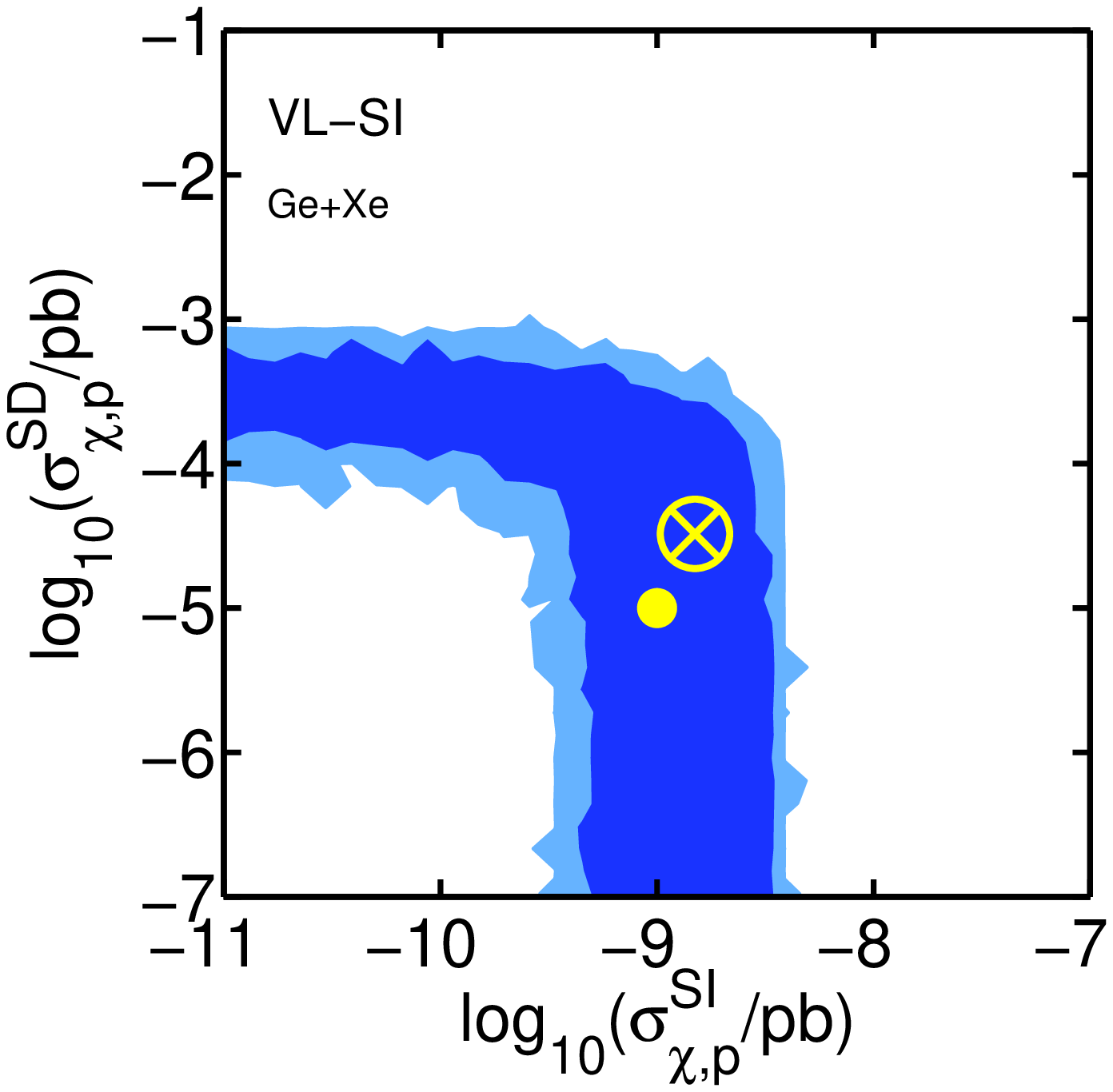}\hspace*{-0.45cm}
\caption{Profile likelihood contours at 68 and 99\% C.L. for Ge+Xe. All uncertainties have been included in the analysis. The yellow dot denotes the benchmark VL-SI and the circled cross the best fit point.}
\label{fig:vlsi}
\end{figure}

\begin{table}
\begin{center}
\begin{tabular}{|c|c|c|c|c|c|}
\hline
 & $\mwimp$ (GeV) & $\sigsi$ (pb) & $\sigsd$ (pb) & $N_{\rm Ge}$ & $N_{\rm Xe}$ \\
\hline
VL-SI & $20$ & $10^{-9}$ & $10^{-5}$ & 40.4\,(39.3) & 65.0\,(61.6)\\
\hline
L-SD & $50$ & $10^{-10}$ & $1.5\times10^{-4}$ & 29.3\,(6.1) & 94.7\,(11.0)\\
\hline
\end{tabular}
\caption{Benchmark points used: VL-SI is a very light WIMP with dominant SI scattering cross-section while L-SD has a significant SD contribution (L-SD). The fifth and sixth columns indicate the total expected recoil events in Ge- and Xe-based experiments for an exposure $\epsilon=300$~kg$\times$yr, in the energy windows [3-100]~keV for Ge and [3-43]~keV for Xe. The number in parenthesis indicates the expected recoils when only SI interactions are considered. In the calculation we considered the mean values of the parameters describing the SDSFs (see Tab.~\ref{tab:SfSD}) and the halo model of Eq.~\ref{eq:halo} with $\locald$=0.4~GeV/cm$^3$, $v_0$=220~km/s, $v_{esc}$=544~km/s and $k$=2.}
\label{tab:bm}
\end{center}
\end{table}
Fig.~\ref{fig:vlsi} shows the 68\% and 99\% confidence level contours for the three 
WIMP parameters projected onto the corresponding two-dimensional plots $\msi$, 
$\msd$, and $\sisd$ for the benchmark VL-SI. The yellow dot represents 
the nominal value and the circled cross is the best-fit point. 
As we showed in Ref.~\citen{cerdeno2013a}, the combination of data from Ge and Xe leads to a substantial reduction in the contours of the reconstructed WIMP parameters. The improved energy threshold also contributes to this. 
In particular, for this benchmark the mass of the
WIMP can be well determined (the contours using only one target would not be closed). However, there remains a 
degeneracy in both cross sections, $\sigsi$ and $\sigsd$, for which only upper 
limits are derived. This is due to the similar sensitivity to SI/SD interactions of Ge and Xe in this point.

\begin{figure}[h]
 \hspace*{-0.5cm}
    \includegraphics[width=0.37\textwidth]{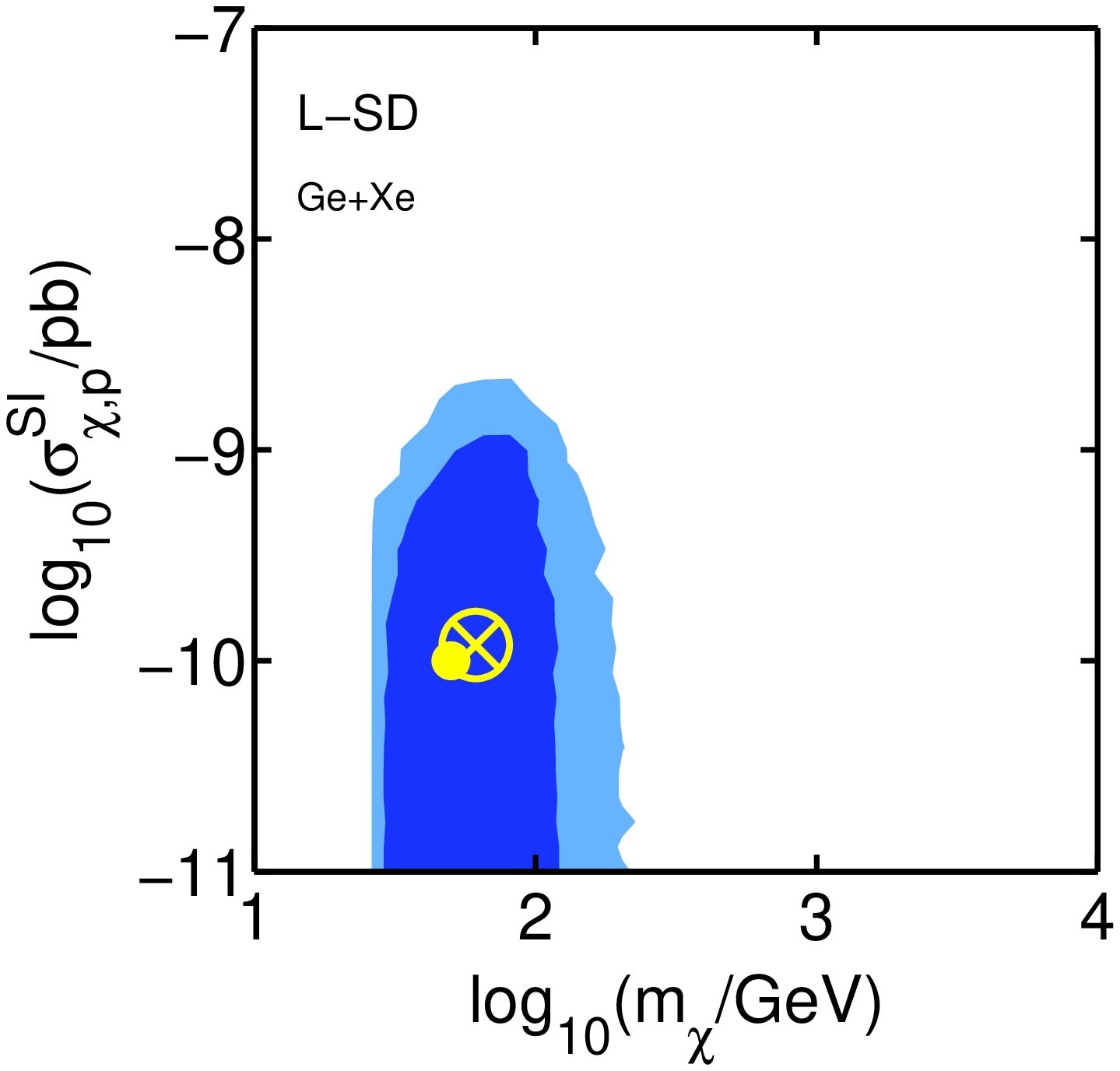}\hspace*{-0.45cm}
    \includegraphics[width=0.37\textwidth]{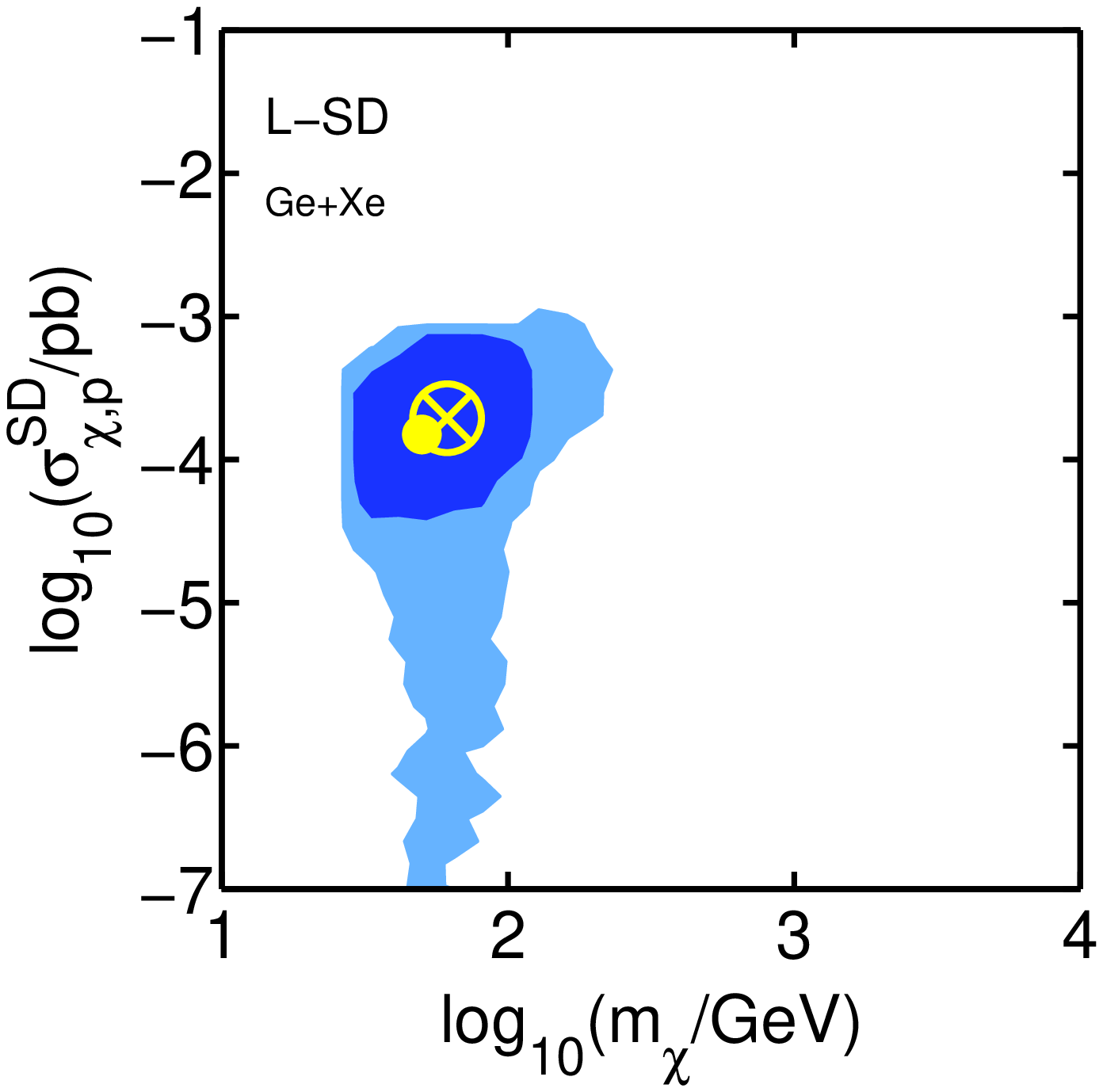}\hspace*{-0.45cm}
    \includegraphics[width=0.37\textwidth]{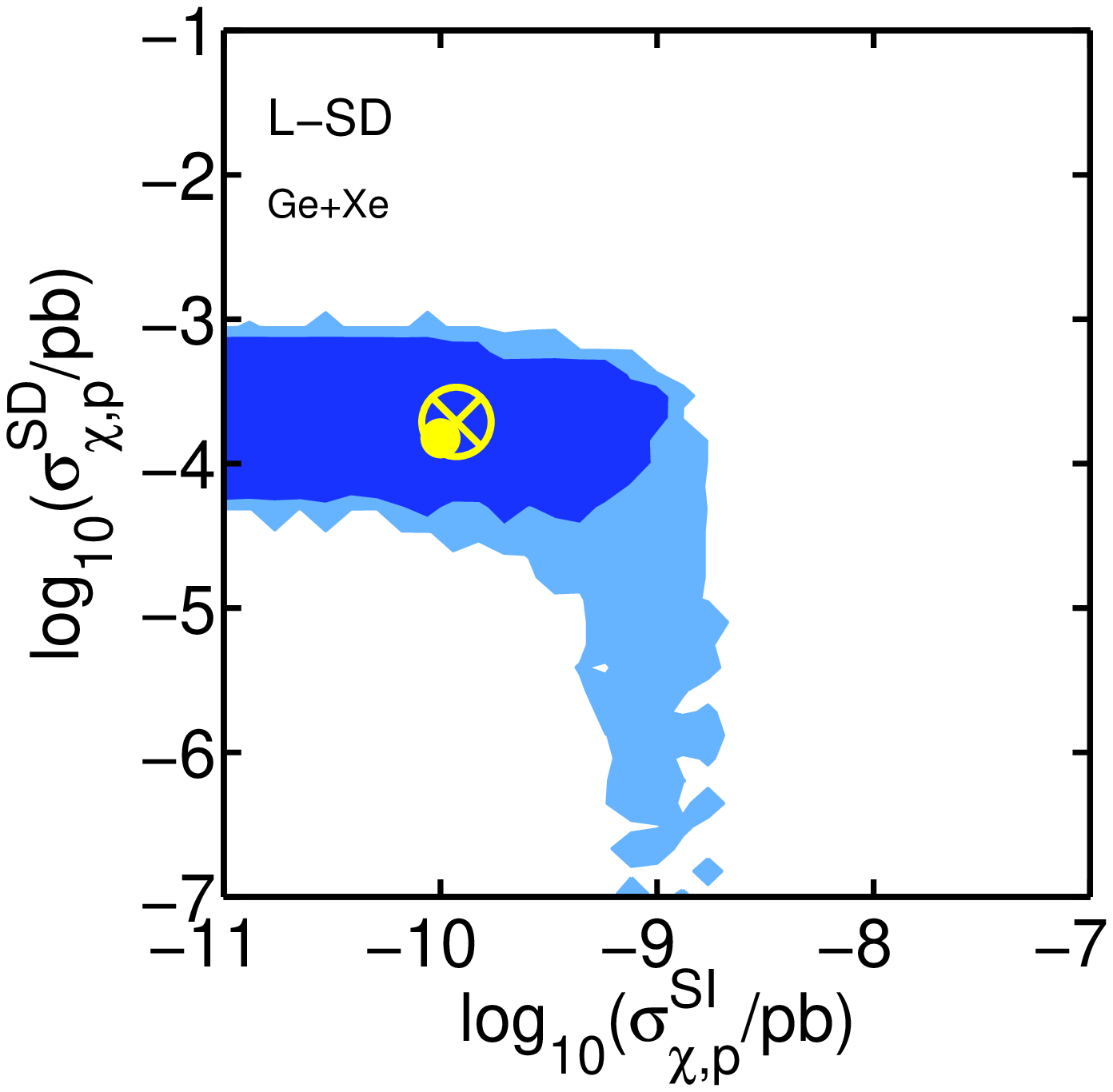}\hspace*{-0.45cm}
\caption{The same as Fig.~\ref{fig:vlsi} but for the L-SD benchmark.}
\label{fig:lsd}
\end{figure}

Analogously, Fig.~\ref{fig:lsd} displays the contour plots for the 
benchmark L-SD. The reconstruction of WIMP parameters is similar, although in this case $\sigsd$ is better
bounded and even a lower limit is derived at 68\%~C.L. Nevertheless, at 99\%~CL the degeneracy between $\sigsi$ and $\sigsd$ still remains and only an upper bound is obtained for $\sigsi$ which is far from the nominal value.

Thus, although the combined data from Ge and Xe experiments can be used to significantly improve the determination of WIMP parameters, the degeneracy in the SI and SD components of the scattering cross section might be difficult to break. As we will argue in the following sections, incorporating data from a third target with a different sensitivity to these components can help solving this problem.


\section{Scintillating bolometers for dark matter searches}
\label{sec:scintBolo}
%
Compared to other hybrid detectors with discrimination capability, 
scintillating bolometers, originally proposed in 1988\cite{GonzalezMestres:1996gb,GonzalezMestres:1988rk}, have the advantage of a wide target choice.
This makes it possible to select intrinsically radiopure materials and combine different 
nuclei to maximize the explored region of the WIMP parameter space 
(high mass number $A$ for large SI coupling, low $A$ to enhance sensitivity to 
light WIMPs, or non-zero nuclear spin for sensitivity to SD interaction, to 
name just a few possibilities).

The energy threshold that has been achieved in the heat signal with cryogenic
detectors is as low as $\sim$1~keV. However, when looking for nuclear recoils, 
the discrimination threshold is determined by the target light yield and the 
sensitivity of the optical detector. Usually this role is played by a second 
low-mass large-area bolometer facing the primary one. Optimizing the 
sensitivity and response of the optical bolometer is a very active ongoing
research field (see for example Refs.\,\citen{isaila2012} and \citen{Beeman2013})
and lower thresholds are expected in a near future.
Nevertheless, in this paper, we follow the same approach of 
Ref.\,\citen{cerdeno2013a} and take a reference energy threshold of 10~keV 
(a value already observed in some scintillating 
bolometers\cite{Amare2006d, Angloher2012}) for the bolometric targets under 
study.

In our previous work\cite{cerdeno2013a} we studied the complementarity of three
scintillating materials: CaWO$_4$, Al$_2$O$_3$ and LiF. CaWO$_4$ is the current 
target of the CRESST experiment\cite{Angloher2012} and was used also 
by the ROSEBUD collaboration in the first underground DM search with light 
and heat discrimination\cite{Cebrian2003}. It constitutes the baseline for 
the EURECA scintillating targets. Al$_2$O$_3$, used by 
ROSEBUD\cite{Amare2005,Calleja2008} and by CRESST in the first phase of the 
experiment\cite{Proebst2002}, is particularly interesting for its sensitivity 
to low mass WIMPs. Finally, LiF, also sensitive to light WIMPs and SD 
interaction, has been used by the ROSEBUD collaboration for DM searches 
and as neutron detector, showing that its use in a DM experiment could allow for thermal 
neutron monitoring\cite{Martinez2012,Coron2010a}. However, the light yields achieved 
so far do not provide a good discrimination threshold, so further 
developments are needed in order to use this target in a DM experiment. 

In this paper we focus on two other targets: CaF$_2$ and NaI.
Fluorine-based scintillators are particularly attractive for DM searches 
because of the sensitivity of $^{19}$F ($J$=1/2, 100\% isotopic abundance)
to SD interactions. Among them, CaF$_2$ presents the highest light 
yield\cite{Kim2010} and has already been used in several DM searches as 
scintillator at room temperature\cite{Ogawa2000,bernabei2002,Shimizu2006}.
It was the target material of the first scintillating bolometer ever 
constructed\cite{Alessandrello1992}, although in that experiment the light 
measurement was performed with a silicon photodiode, less sensitive than 
the semiconductor bolometers usually used in recent setups\cite{Bobin1997}.
Scintillation at low temperature has been studied for pure and europium-activated targets,
resulting in good scintillation at 1~K specially 
for doped samples\cite{Coron2004,Mikhailik2006}, although the radiopurity 
levels achieved in this case are usually worse.
\\
NaI, on the other hand, is one of the most widely used scintillators for
$\gamma$ spectroscopy due to its very high light yield.
As mentioned above, this is the target used by DAMA/LIBRA and other proposed
DM experiments looking for annual modulation\cite{Amare2012,Cherwinka:2011ij}.
Although NaI is usually doped with Tl for room temperature applications, the
pure material is known to scintillate better at
temperatures of a few Kelvin\cite{vanSciver58} (nevertheless, an increase in light yield of the
Tl-doped material below 30~K has been recently
reported\cite{Sailer:2012ua,Coron2013}). Despite its high light yield at low
temperature and intrinsic interest for DM searches, this material has not
been tested yet as a bolometer due to its fragility and high hygroscopicity.

\section{Results with bolometric targets}
\label{sec:results}
\begin{table}
\begin{center}
\begin{tabular}{|c|c|c|c|c|c|c|c|}
\hline
 &  $\mwimp$  &  $\sigsi$ &  $\sigsd$ & $N_{\rm NaI}$ & $N_{\rm NaI}$ & $N_{\rm NaI}$ & $N_{\rm CaF_2}$ \\
          &   (GeV) &  (pb) &  (pb) & $q$=0.85 & $q$=1 & $q$=1.15 & $q$=1 \\
\hline
VL-SI & $20$ & $10^{-9}$  & $10^{-5}$          & 3.5\,(2.9)  & 6.3\,(5.3) & 9.5\,(8.2)  & 22.2\,(4.3) \\
\hline
L-SD  & $50$ & $10^{-10}$ & $1.5\times10^{-4}$ & 51.2\,(3.7) & 60.9\,(4.5) & 69.2\,(5.2) & 364.2\,(0.9) \\
\hline
\end{tabular}
\caption{Number of WIMP recoils expected in the bolometric targets for the benchmarks (BM) described in Sec.~\ref{sec:wimpParam}.  In both cases data correspond to an exposure of $\epsilon=300$~kg$\times$yr and [10-100]~keV energy window.  The number in parenthesis indicate the contribution from SI interaction.  For NaI three different values of the quenching factor have been considered.  }
\label{tab:bmScint}
\end{center}
\end{table}

\begin{figure}[h]
 \hspace*{-0.5cm}
    \includegraphics[width=0.37\textwidth]{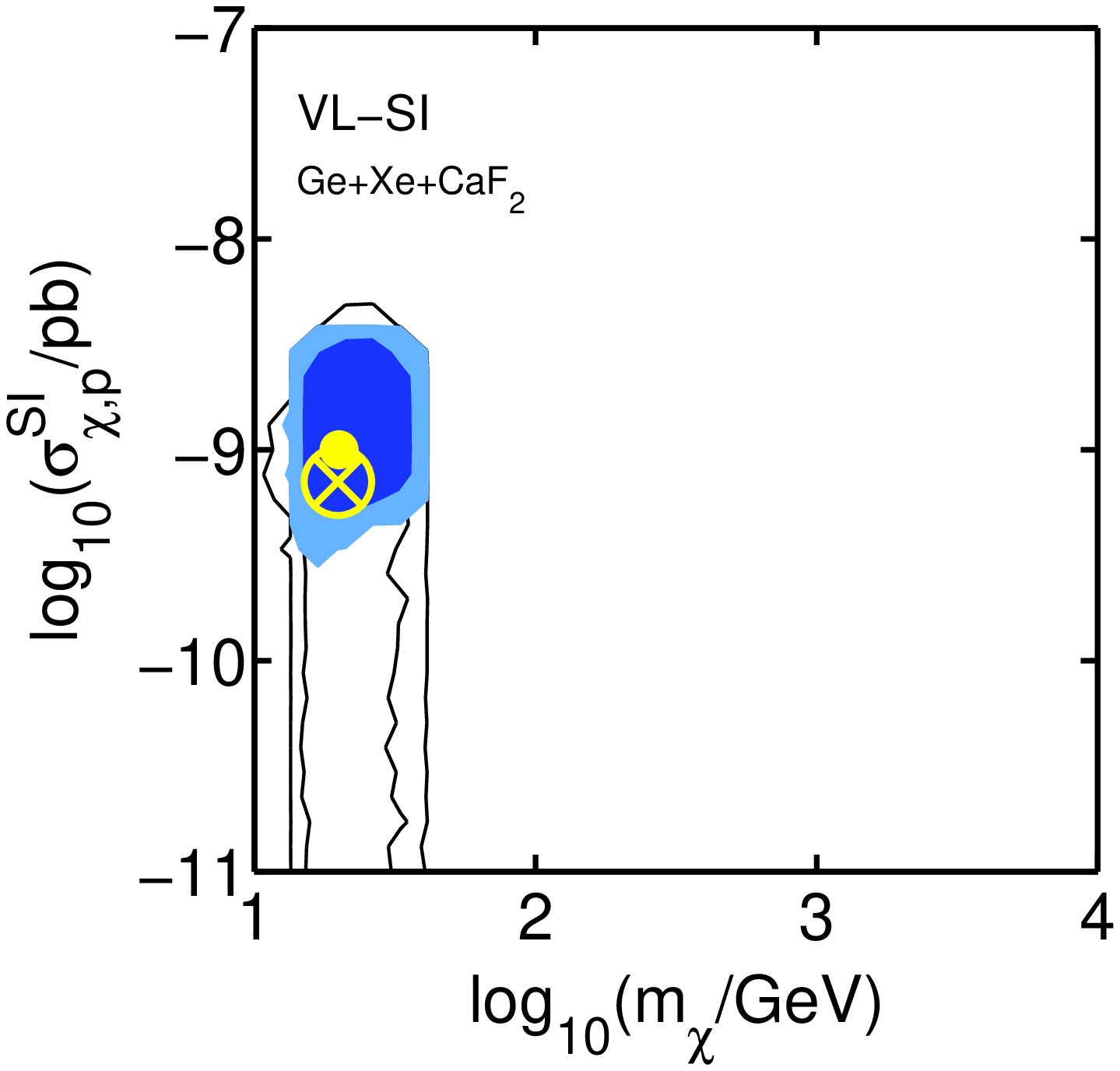}\hspace*{-0.45cm}
    \includegraphics[width=0.37\textwidth]{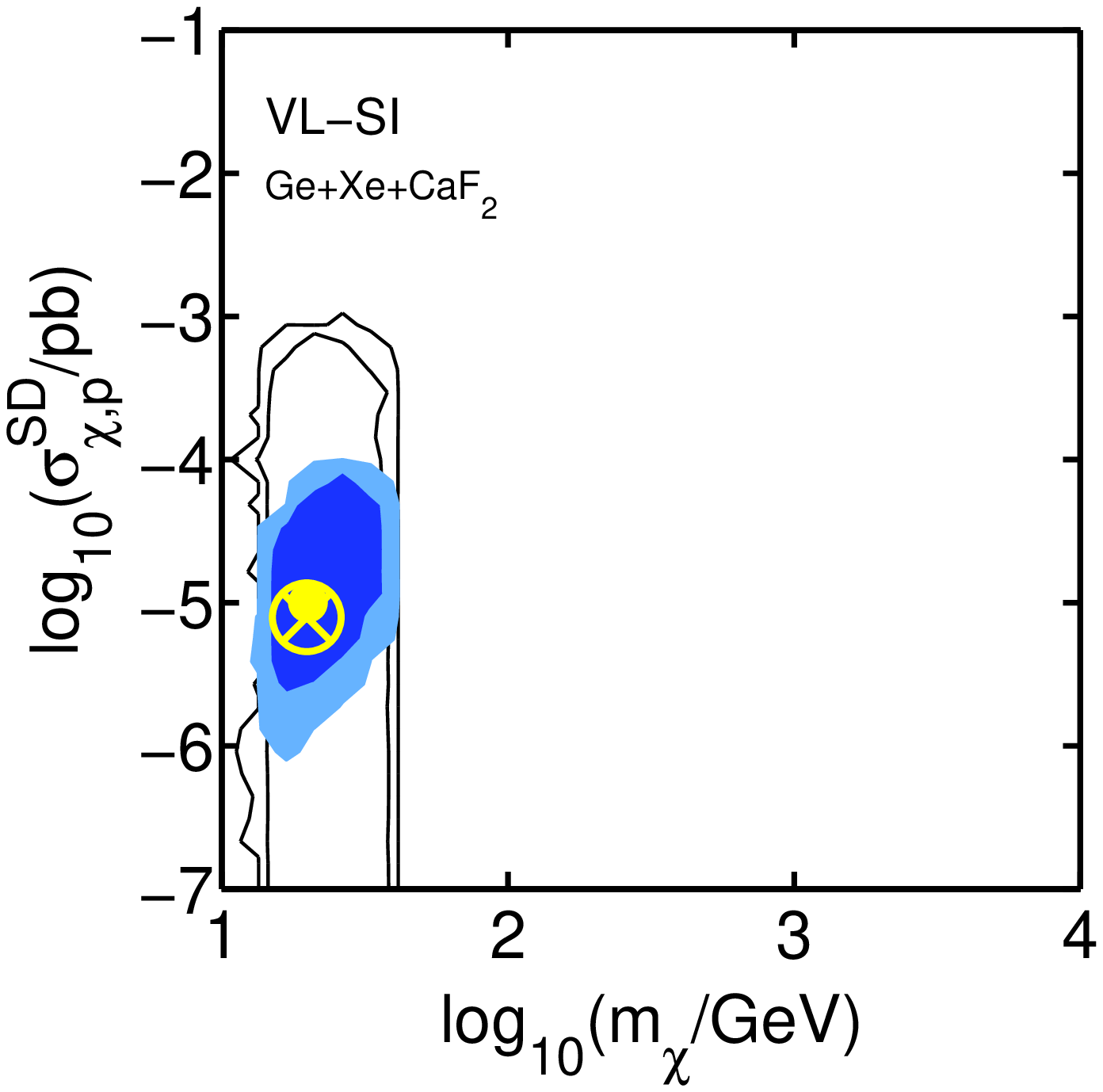}\hspace*{-0.45cm}
    \includegraphics[width=0.37\textwidth]{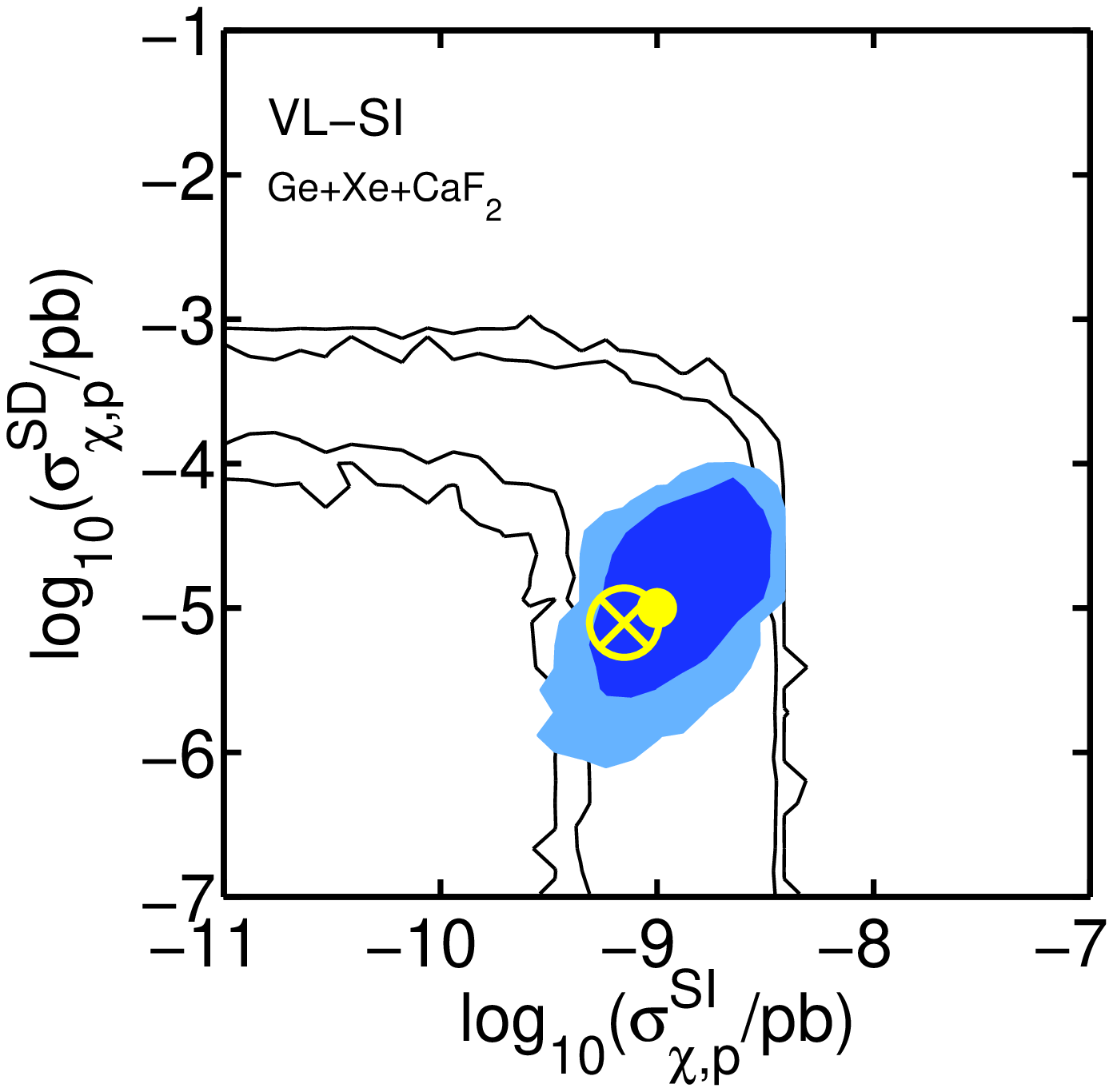}\hspace*{-0.45cm}
\caption{Profile likelihood contours at 68 and 99\% C.L. The blue contours correspond to
Ge+Xe+CaF$_2$ while the empty one to only Ge+Xe. 
The yellow dot denotes the benchmark VL-SI and the circled cross
the best fit point.}
\label{fig:vlsiCaF2}
\end{figure}

\begin{figure}[t!]
 \hspace*{-1cm}
    \includegraphics[width=0.37\textwidth]{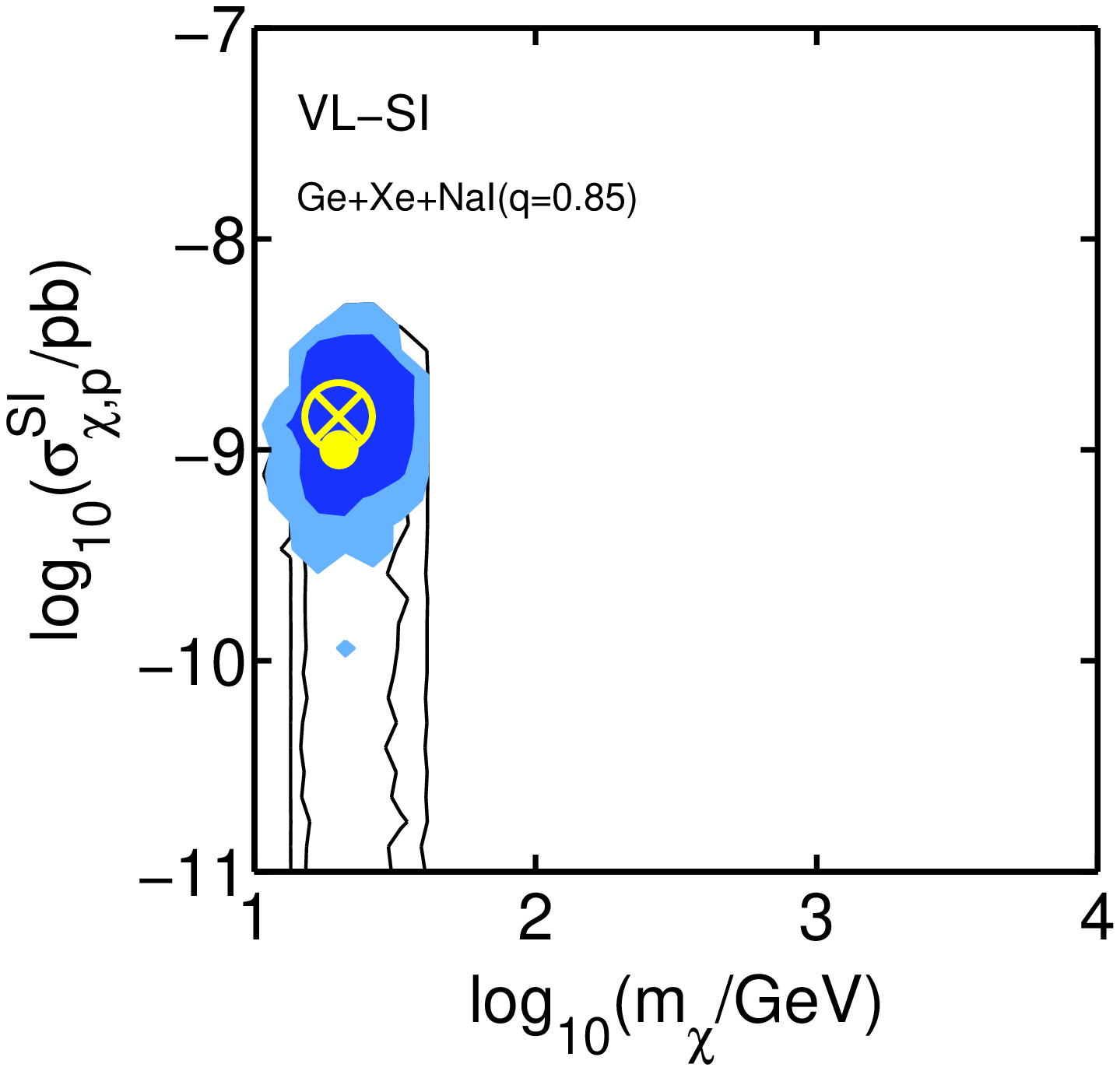}\hspace*{-0.45cm}
    \includegraphics[width=0.37\textwidth]{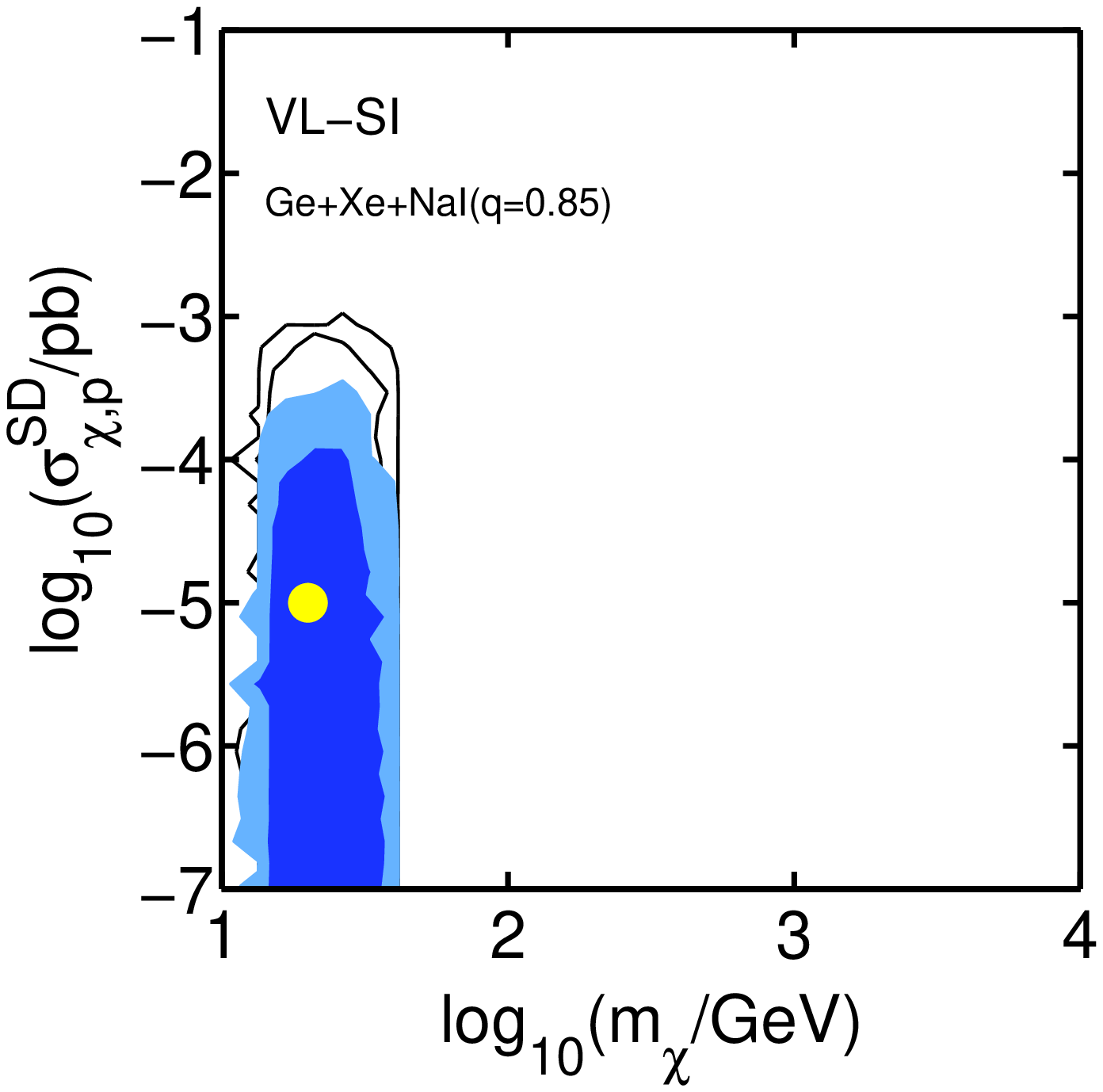}\hspace*{-0.45cm}
    \includegraphics[width=0.37\textwidth]{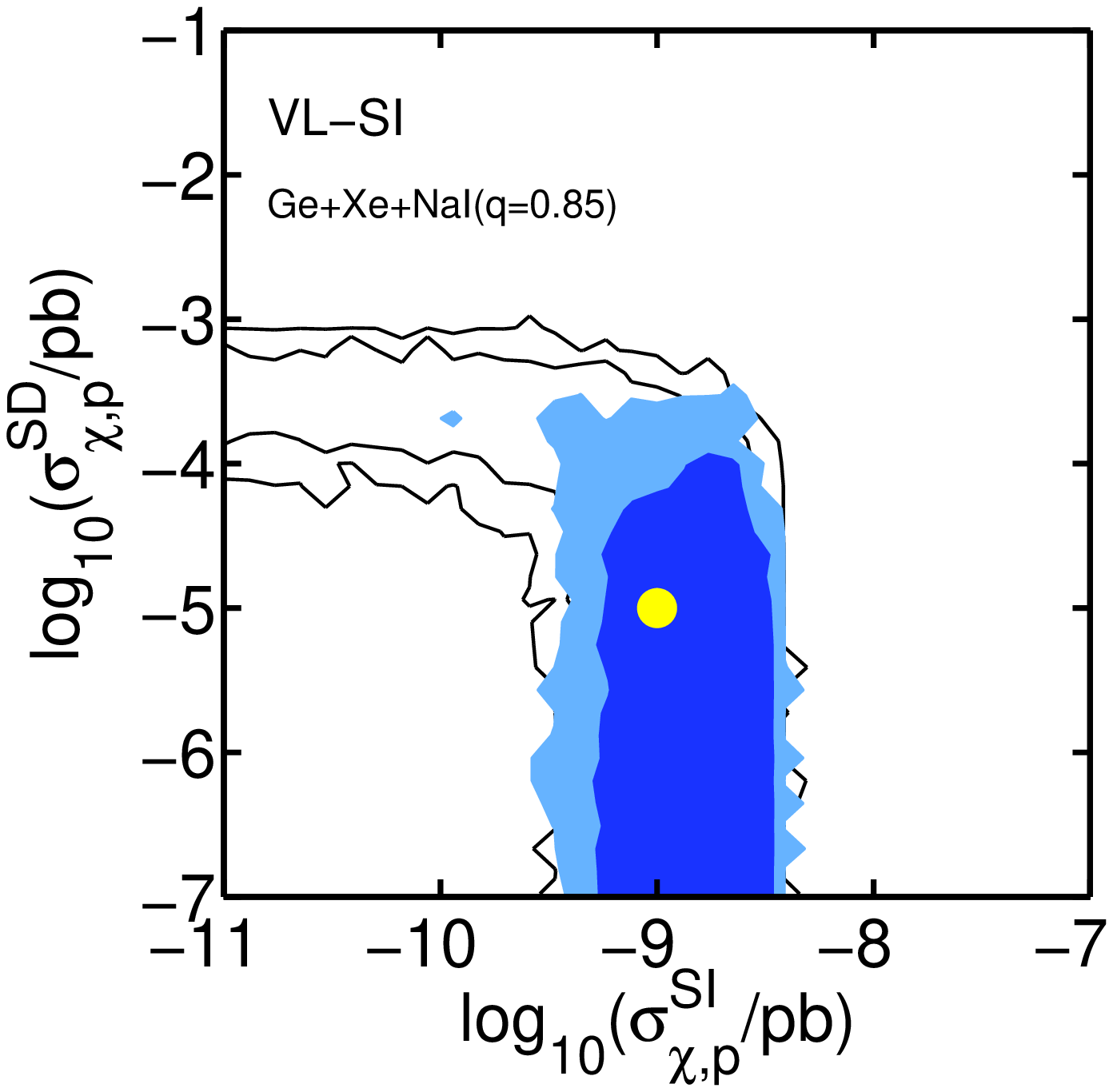}\hspace*{-0.45cm}
  \\[-0.5cm]
 \hspace*{-1cm}
    \includegraphics[width=0.37\textwidth]{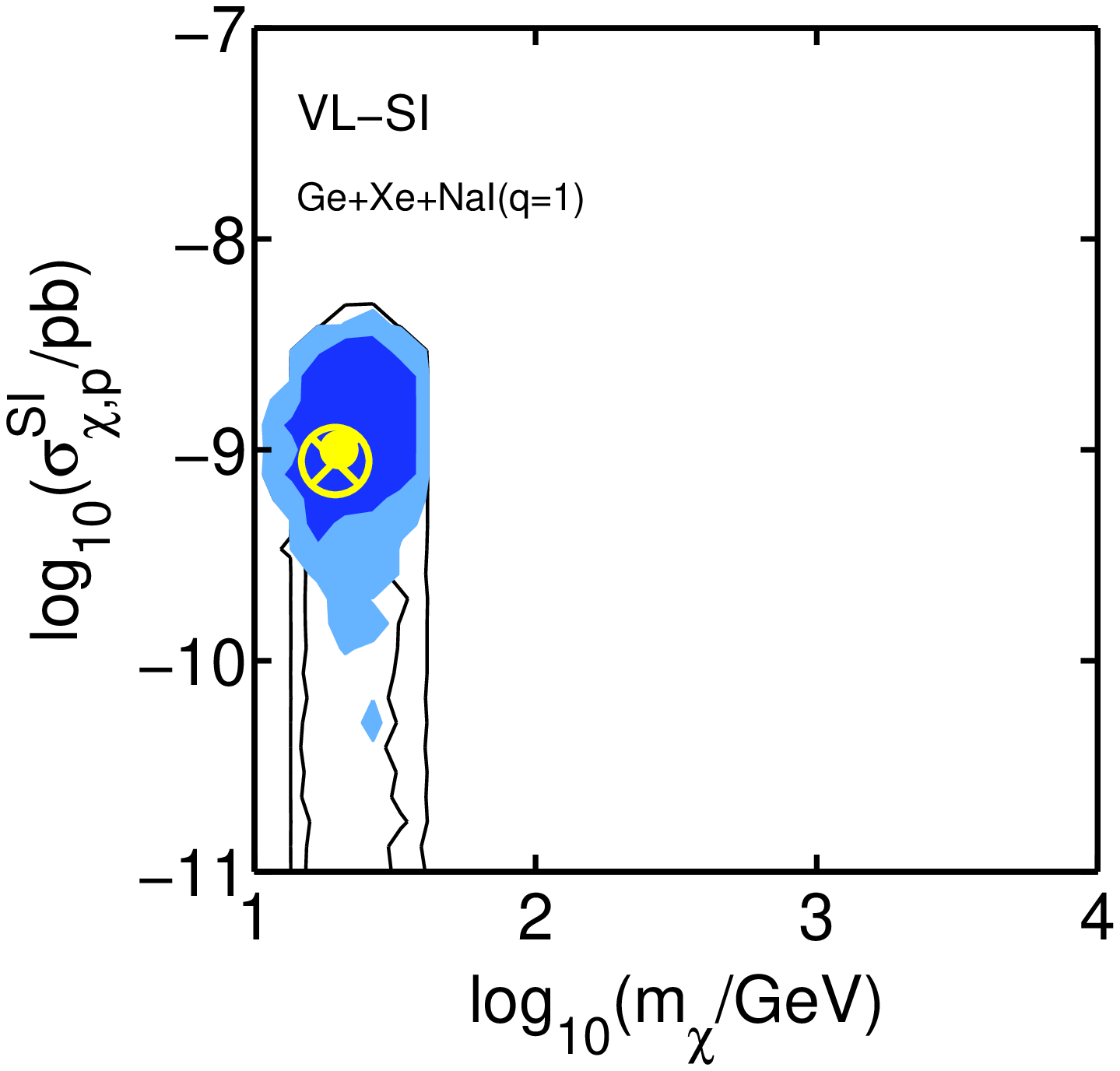}\hspace*{-0.45cm}
    \includegraphics[width=0.37\textwidth]{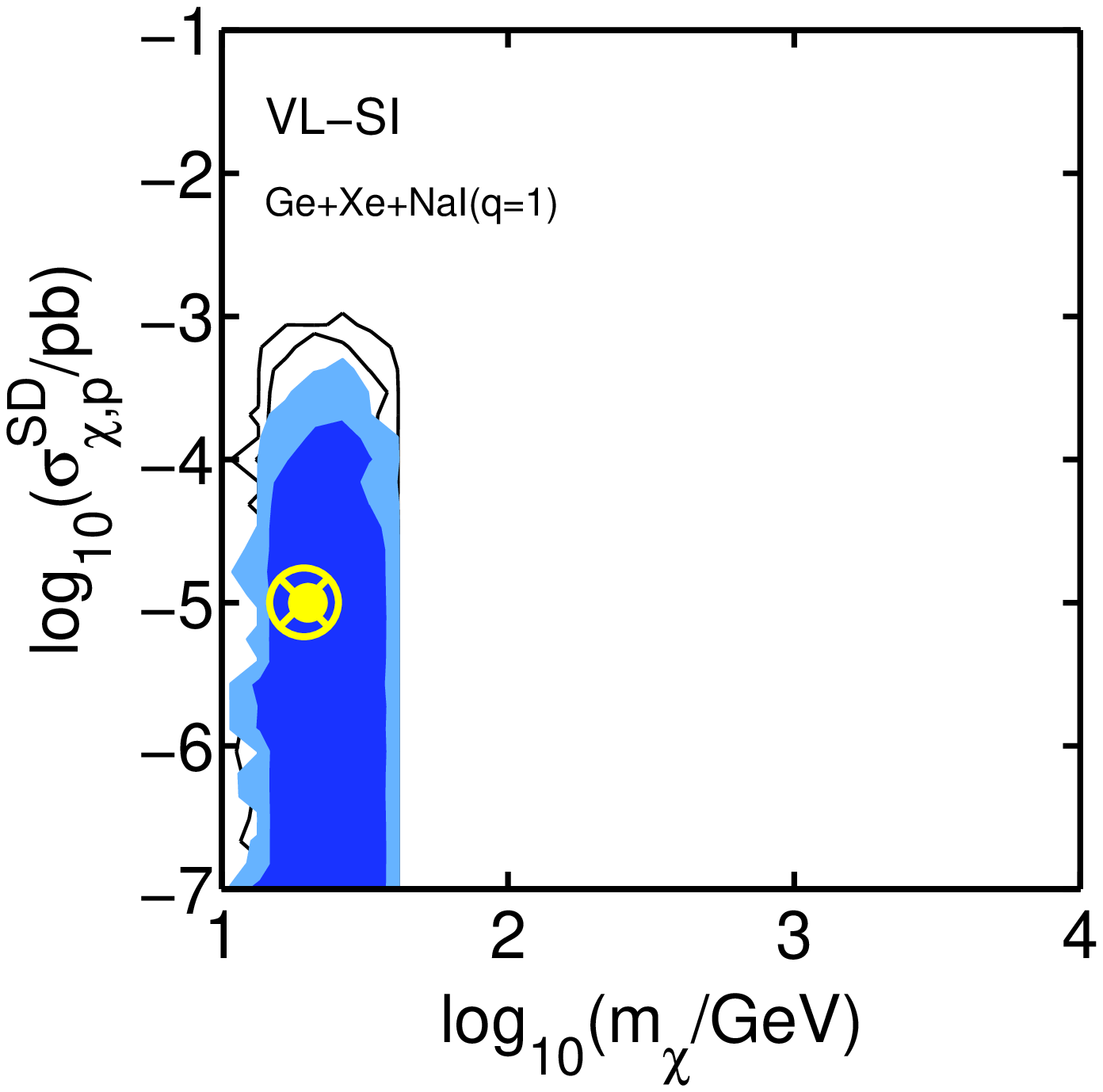}\hspace*{-0.45cm}
    \includegraphics[width=0.37\textwidth]{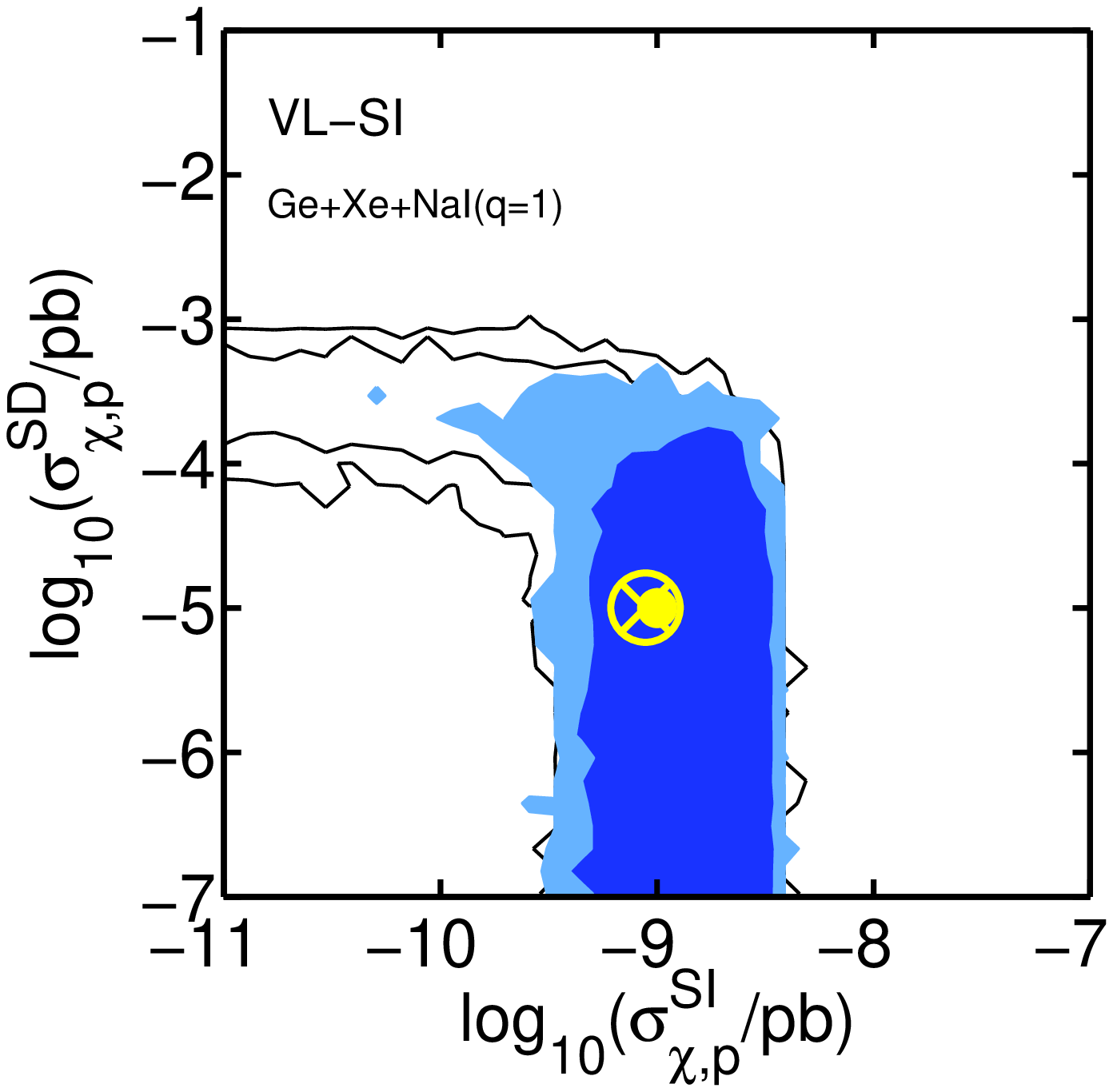}\hspace*{-0.45cm}
  \\[-0.5cm]
 \hspace*{-1cm}
    \includegraphics[width=0.37\textwidth]{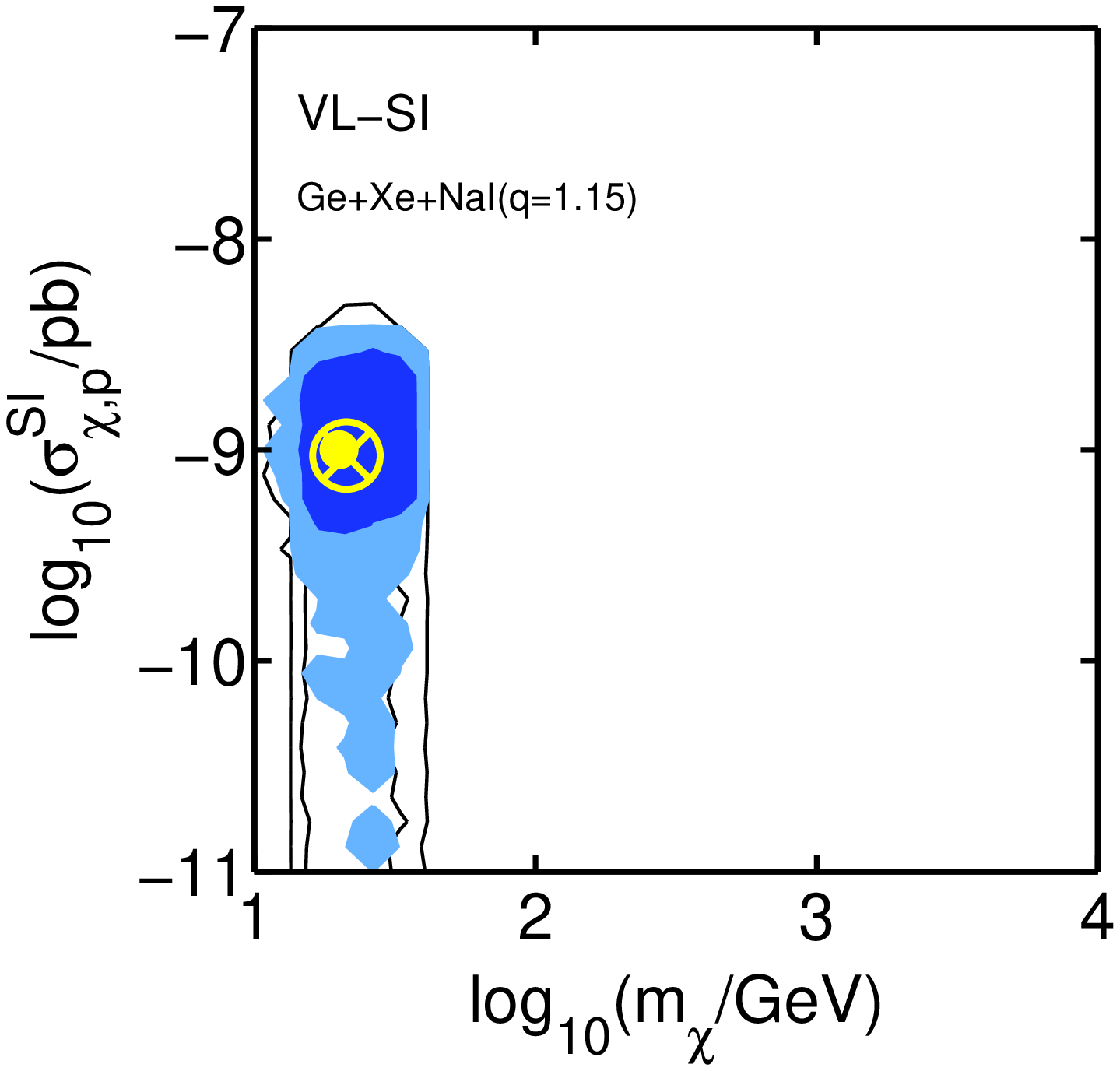}\hspace*{-0.45cm}
    \includegraphics[width=0.37\textwidth]{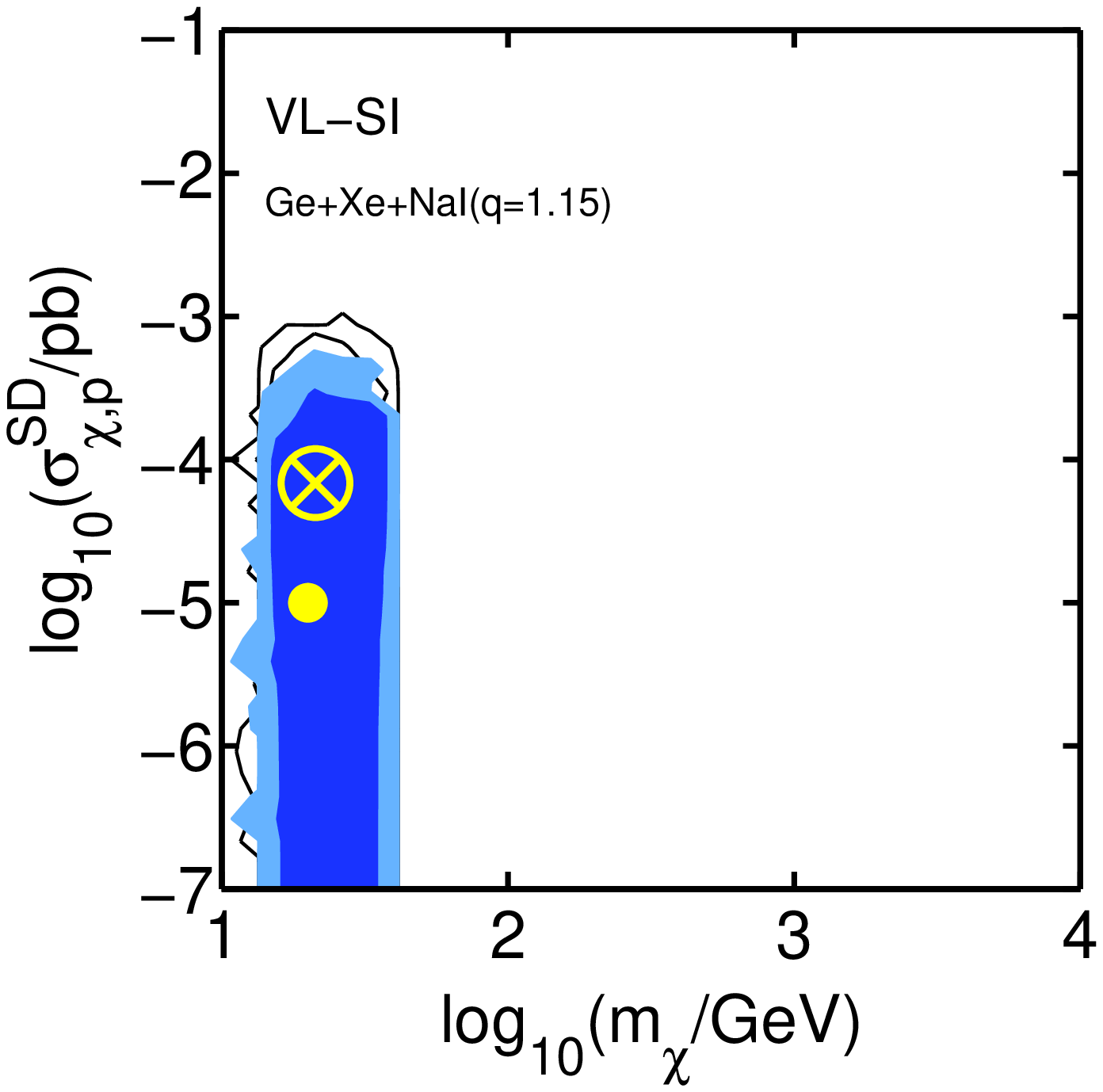}\hspace*{-0.45cm}
    \includegraphics[width=0.37\textwidth]{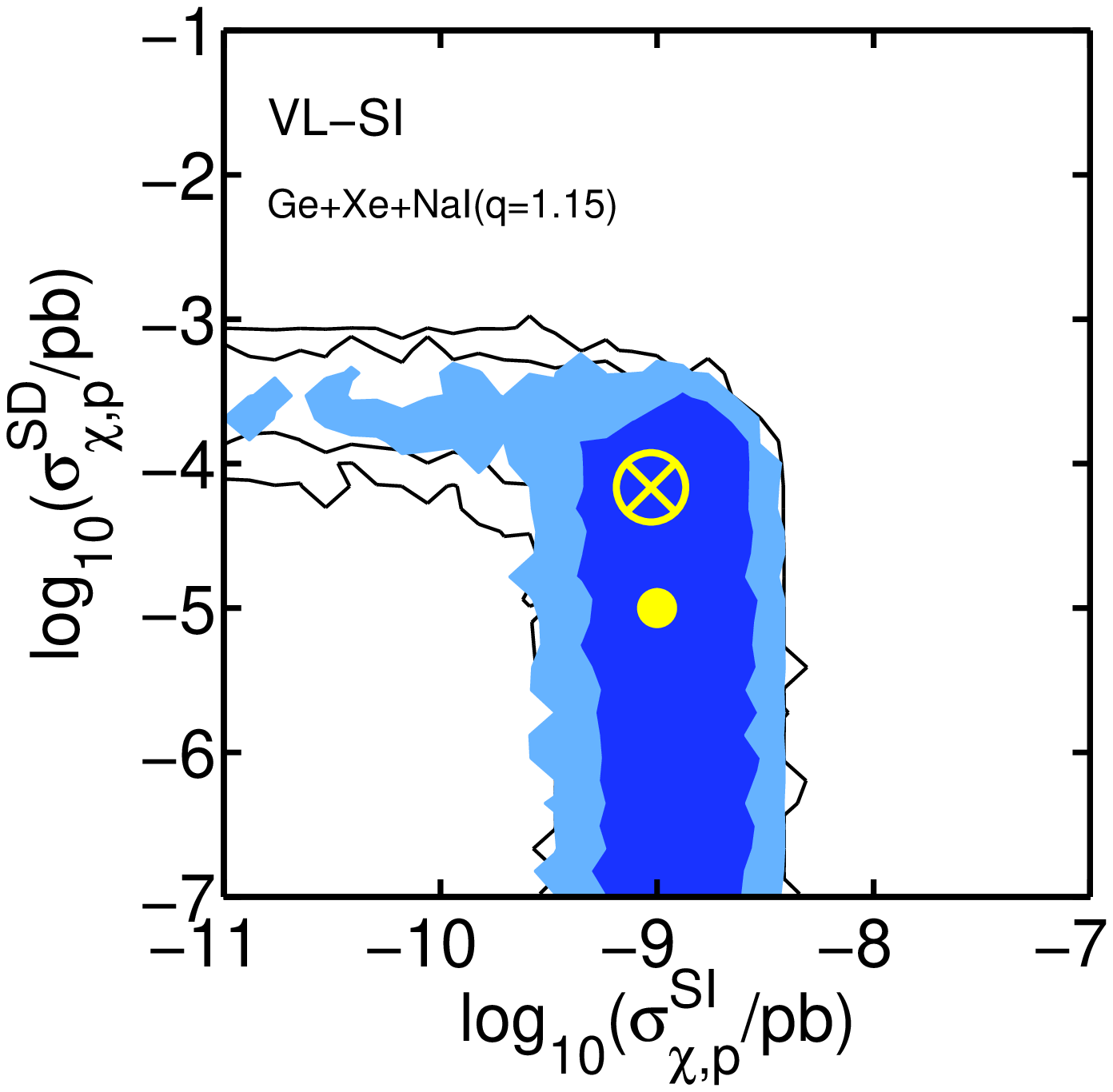}\hspace*{-0.45cm}
\caption{The same as Fig.~\ref{fig:vlsiCaF2} but for the combination Ge+Xe+NaI. From top to bottom the
quenching factor for NaI is 0.85, 1 and 1.15, respectively.}
\label{fig:vlsiNaI}
\end{figure}

\begin{figure}[h]
 \hspace*{-1cm}
    \includegraphics[width=0.37\textwidth]{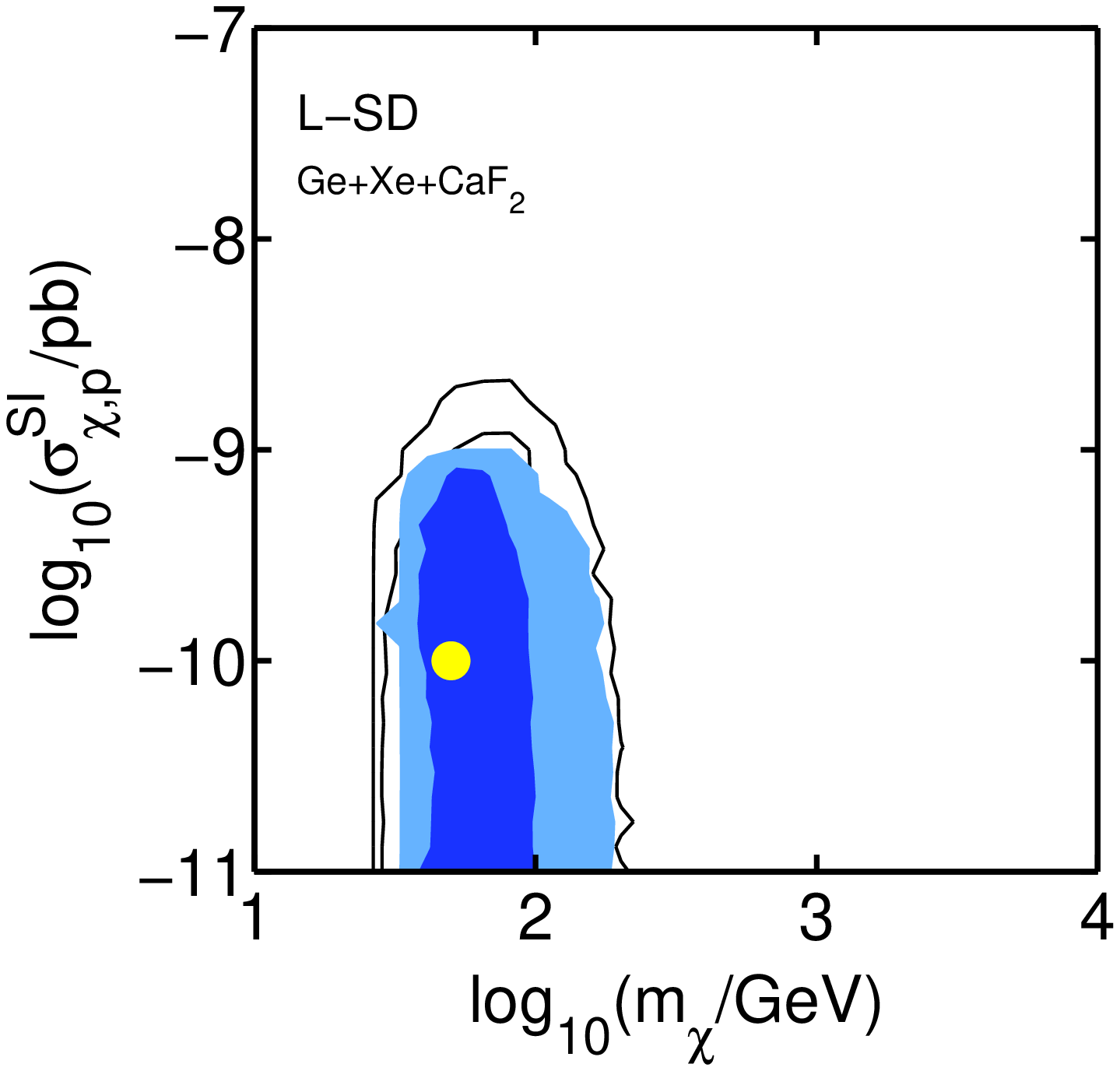}\hspace*{-0.45cm}
    \includegraphics[width=0.37\textwidth]{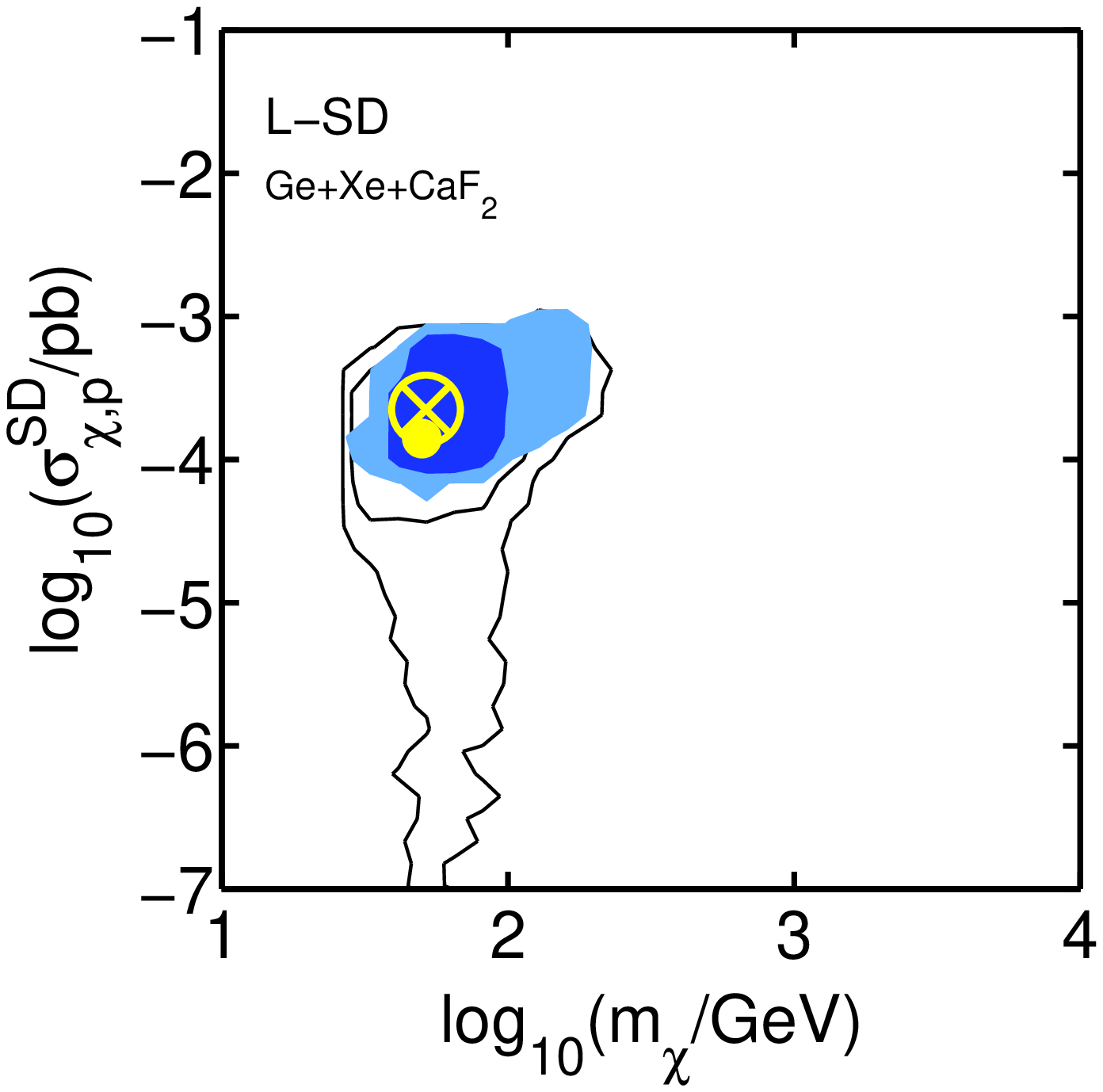}\hspace*{-0.45cm}
    \includegraphics[width=0.37\textwidth]{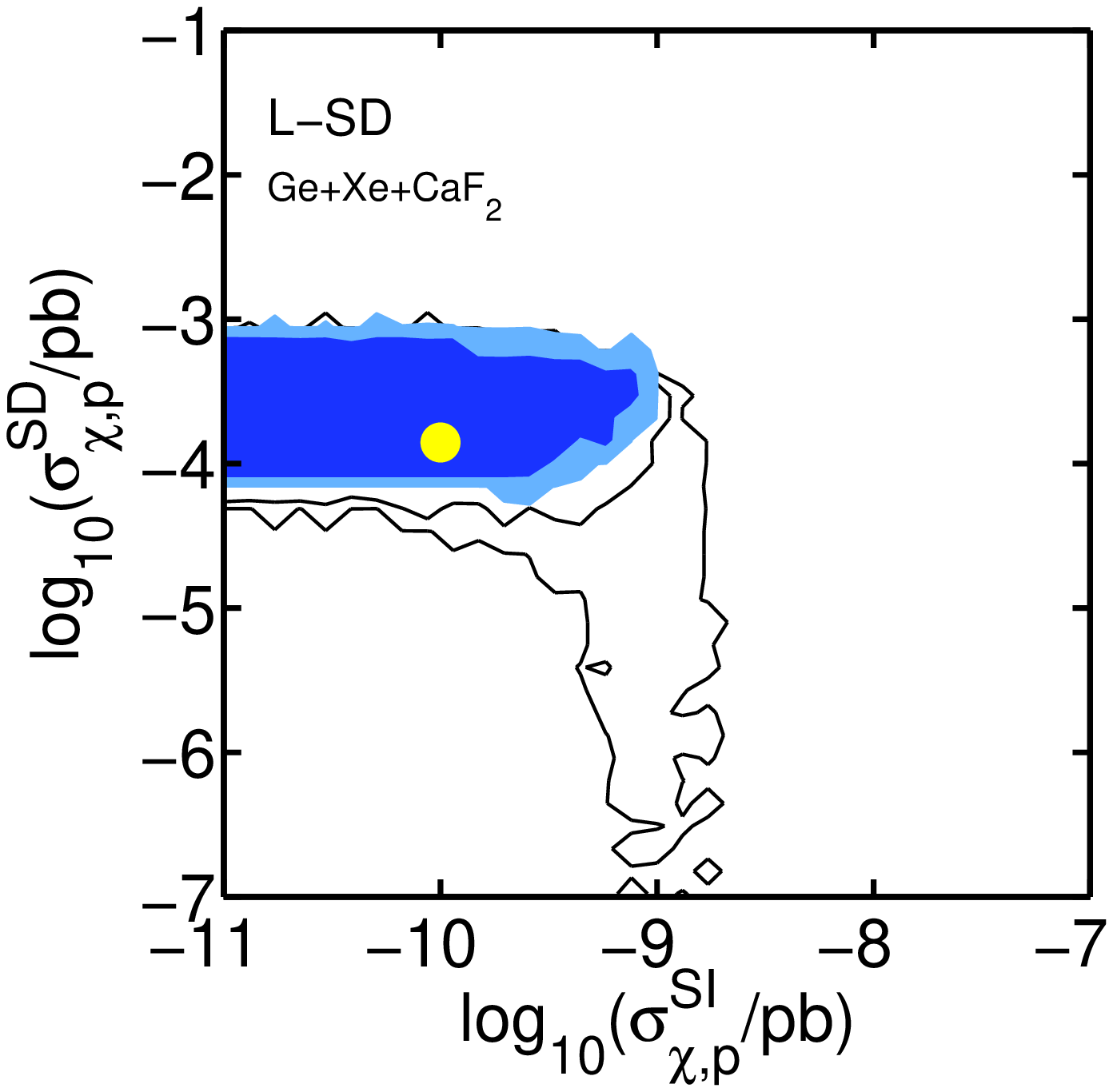}\hspace*{-0.45cm}
\caption{The same as Fig.~\ref{fig:vlsiCaF2} but for the L-SD benchmark.}
\label{fig:lsdCaF2}
\end{figure}

\begin{figure}[t!]
 \hspace*{-1cm}
    \includegraphics[width=0.37\textwidth]{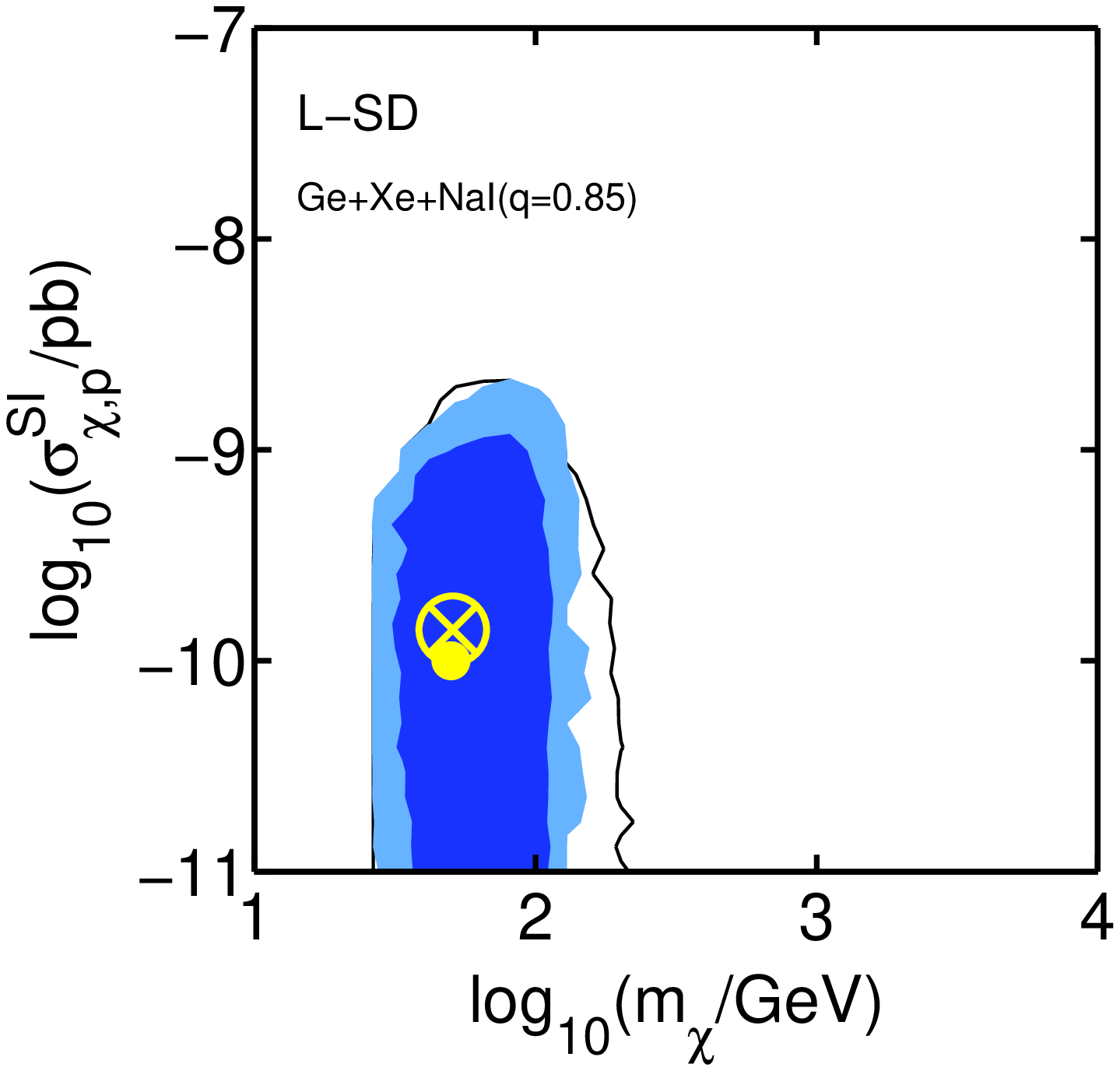}\hspace*{-0.45cm}
    \includegraphics[width=0.37\textwidth]{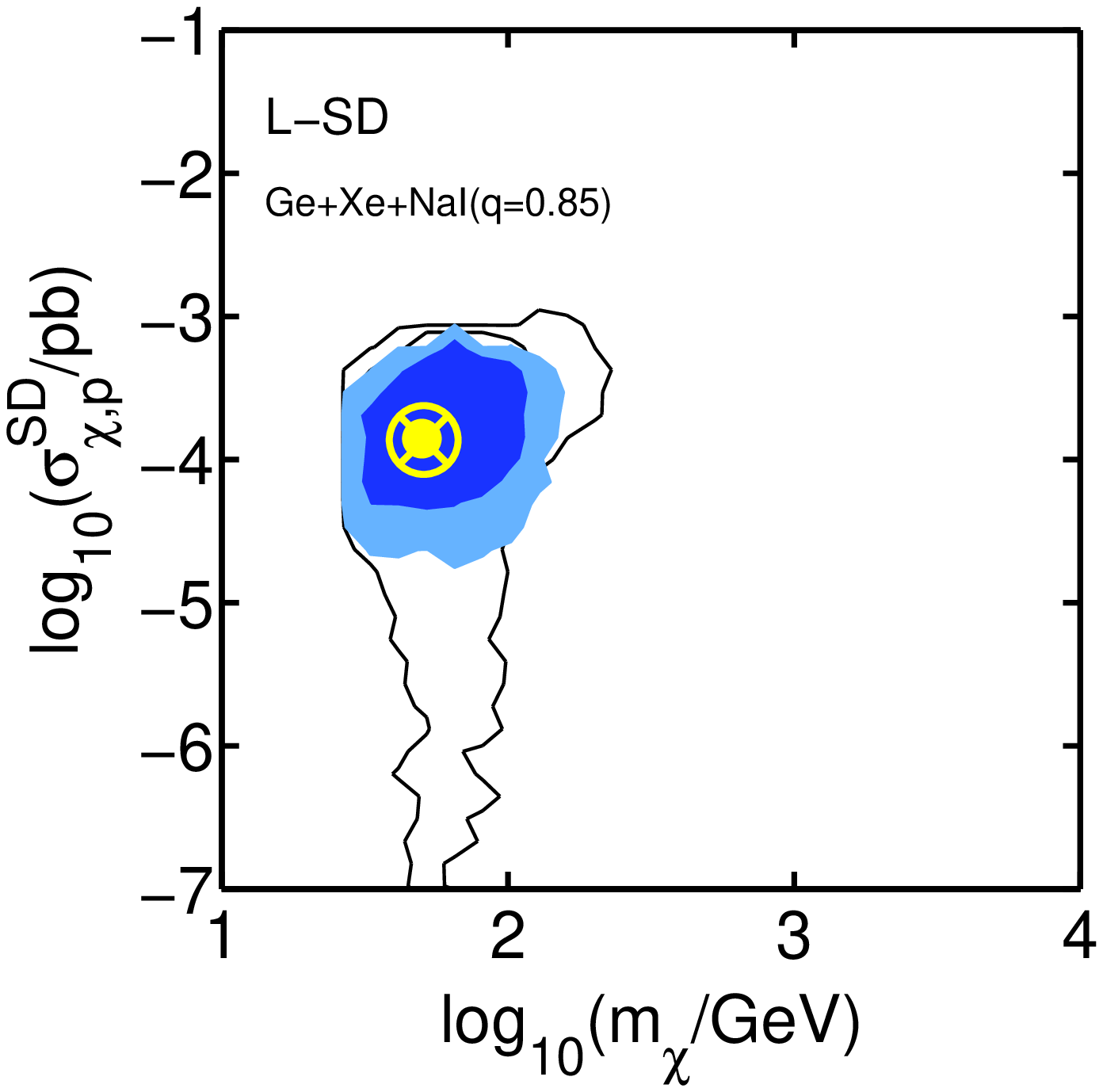}\hspace*{-0.45cm}
    \includegraphics[width=0.37\textwidth]{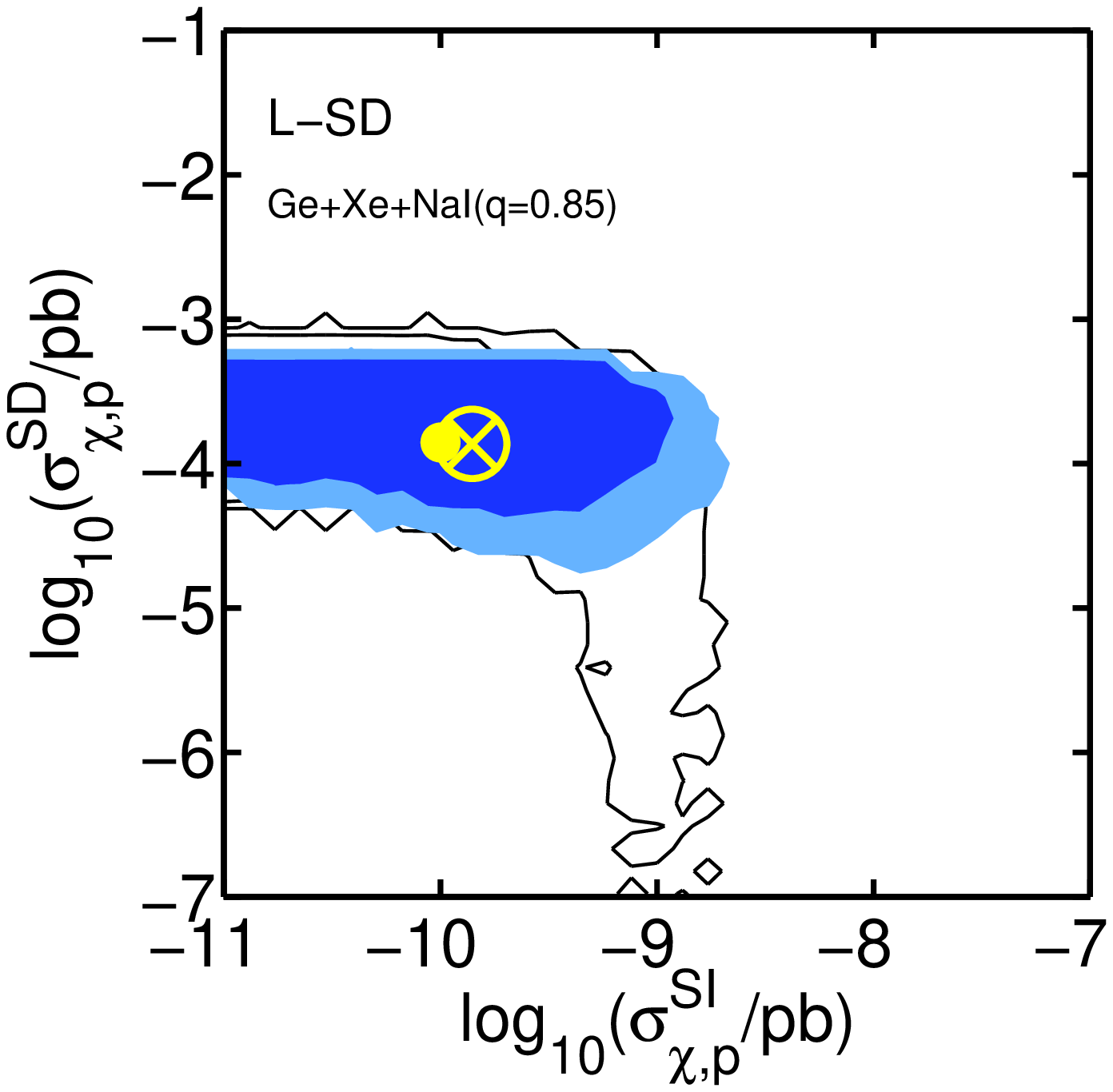}\hspace*{-0.45cm}
  \\[-0.5cm]
 \hspace*{-1cm}
    \includegraphics[width=0.37\textwidth]{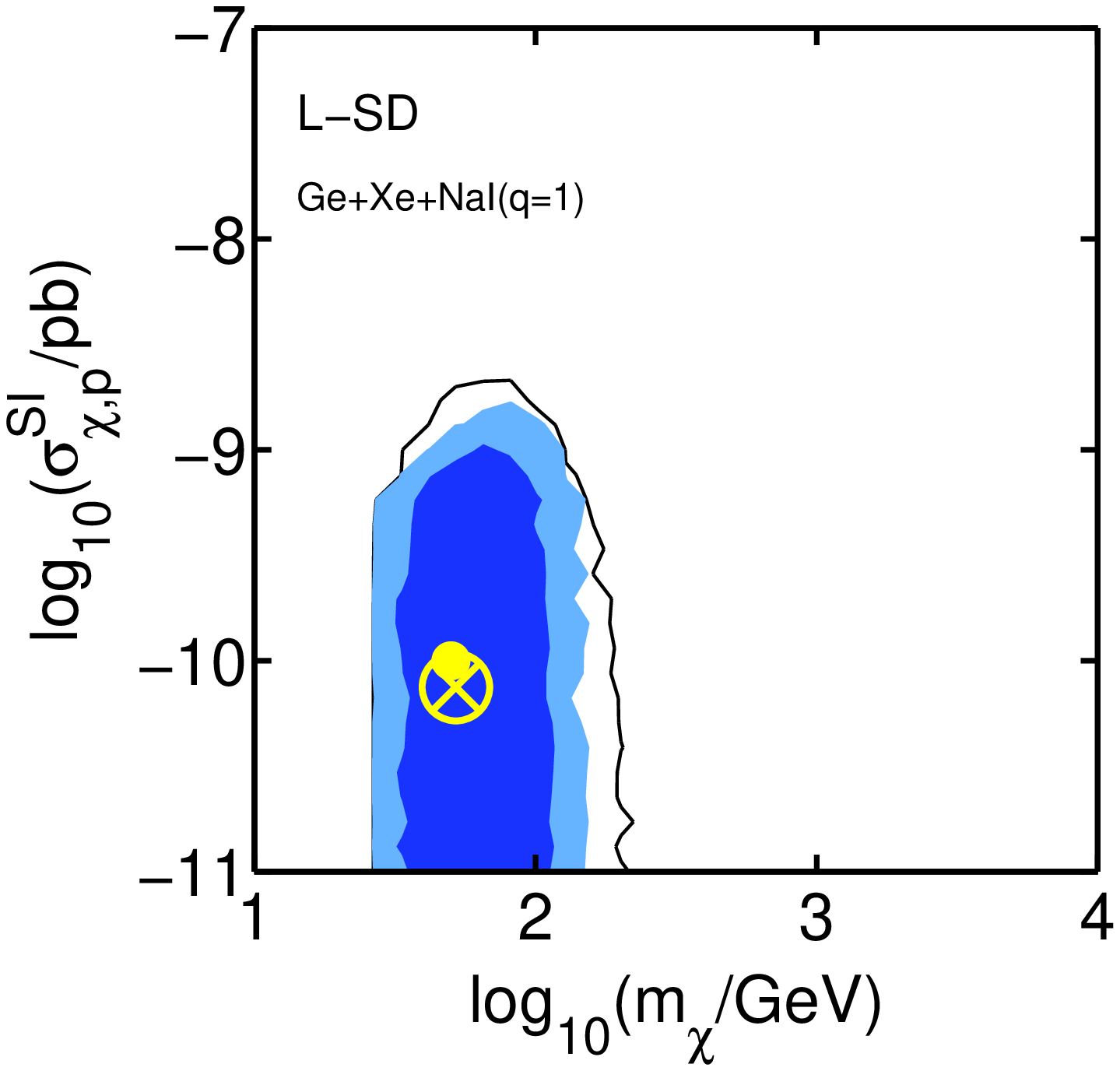}\hspace*{-0.45cm}
    \includegraphics[width=0.37\textwidth]{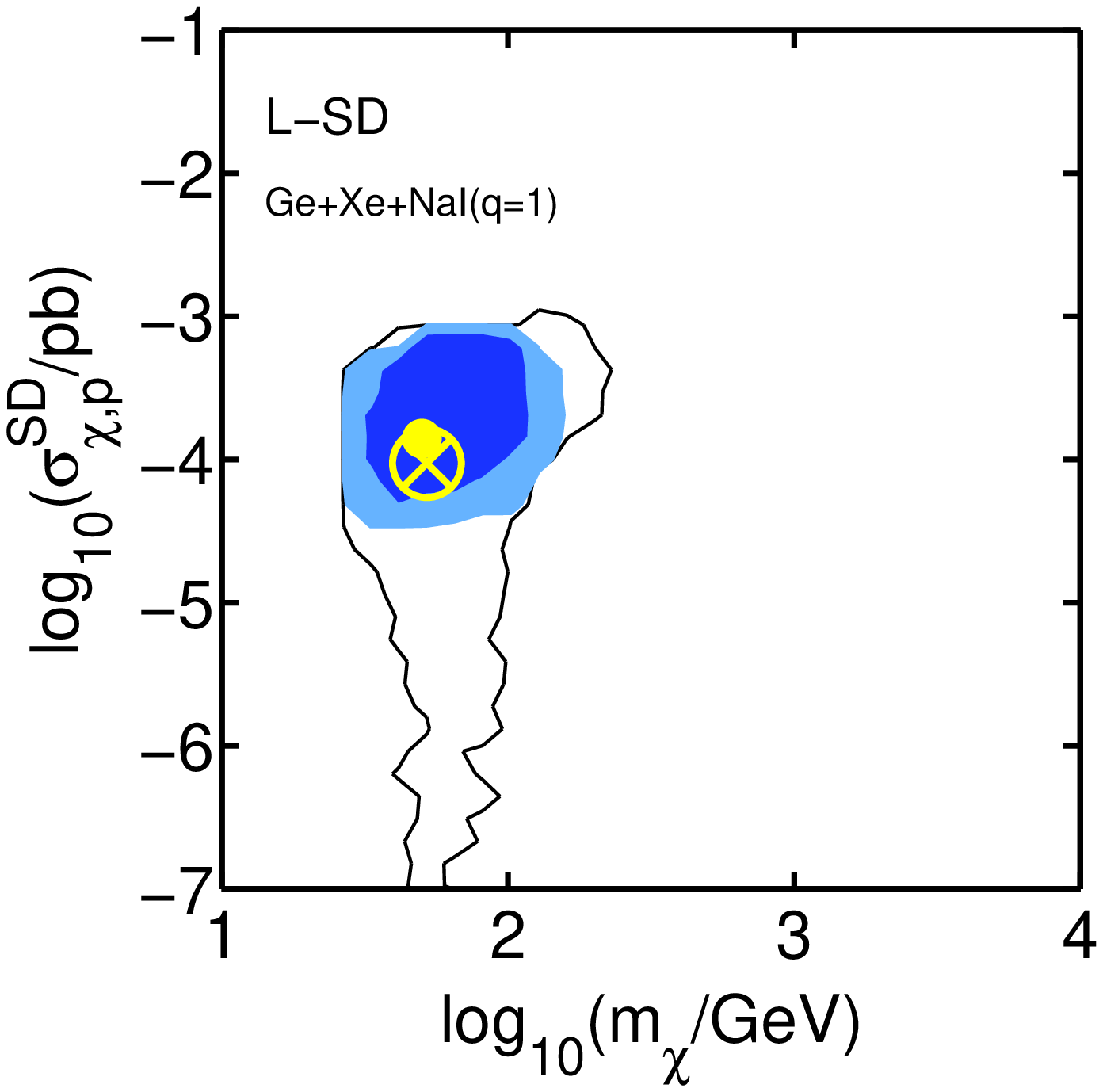}\hspace*{-0.45cm}
    \includegraphics[width=0.37\textwidth]{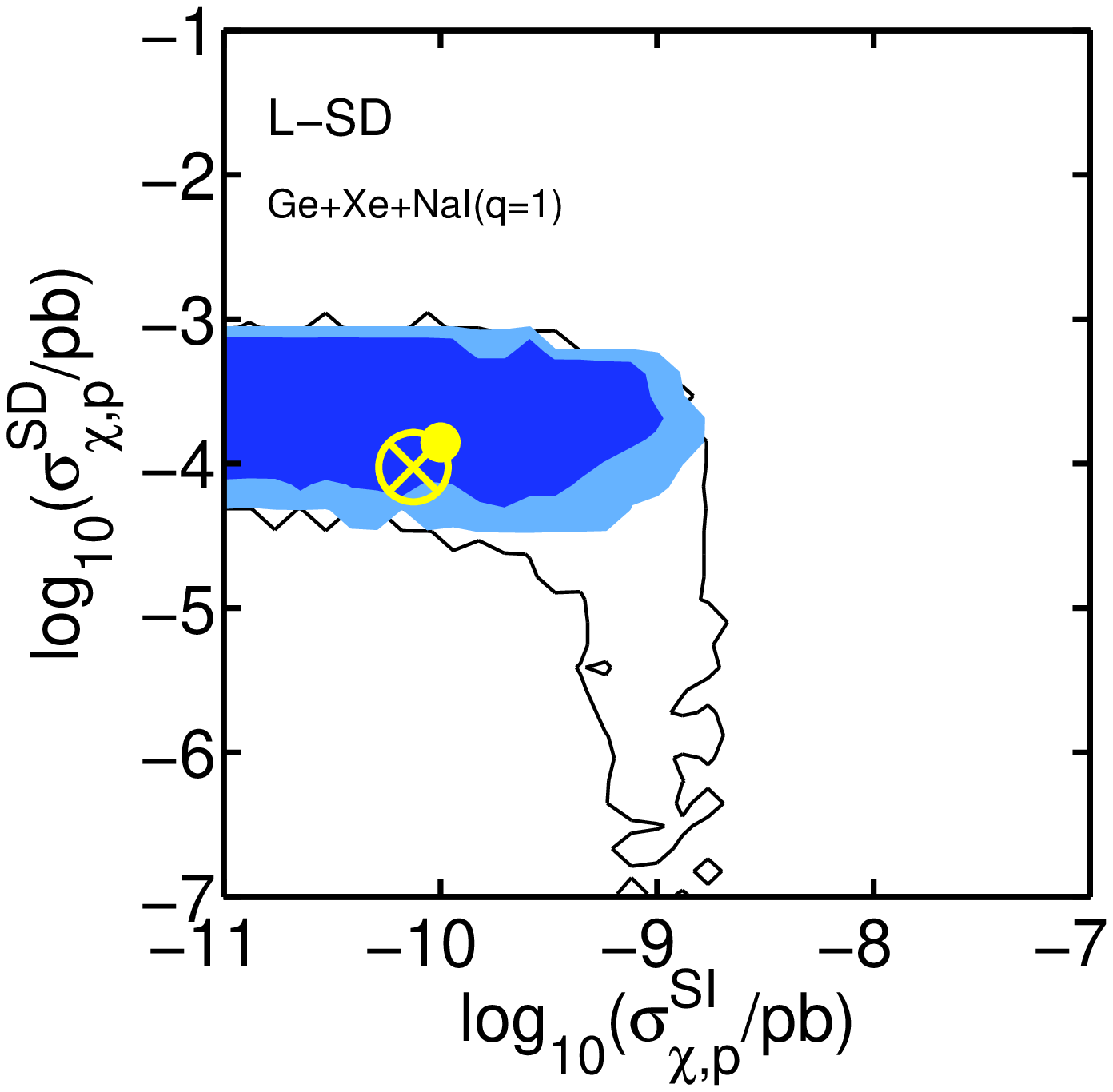}\hspace*{-0.45cm}
  \\[-0.5cm]
 \hspace*{-1cm}
    \includegraphics[width=0.37\textwidth]{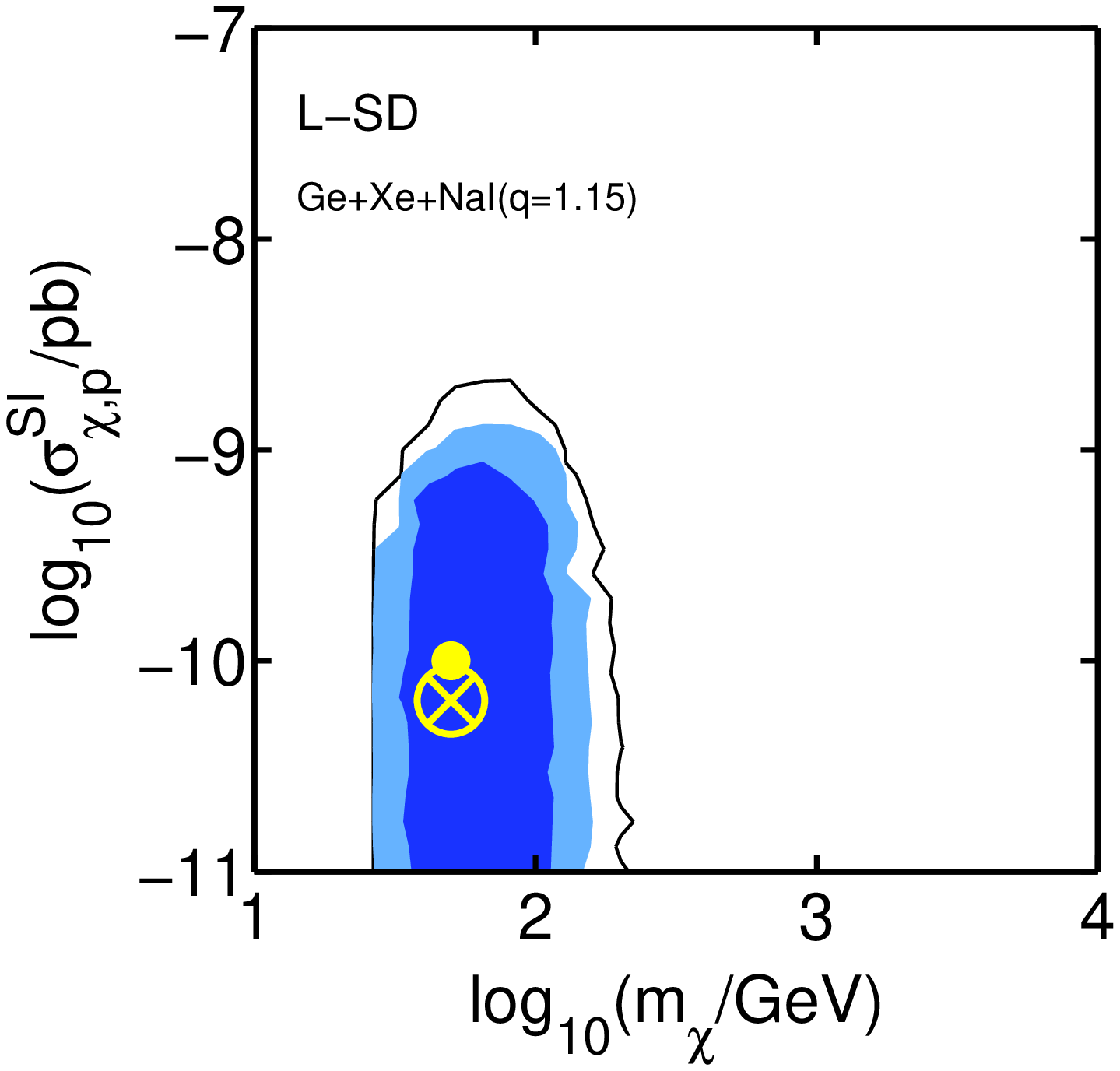}\hspace*{-0.45cm}
    \includegraphics[width=0.37\textwidth]{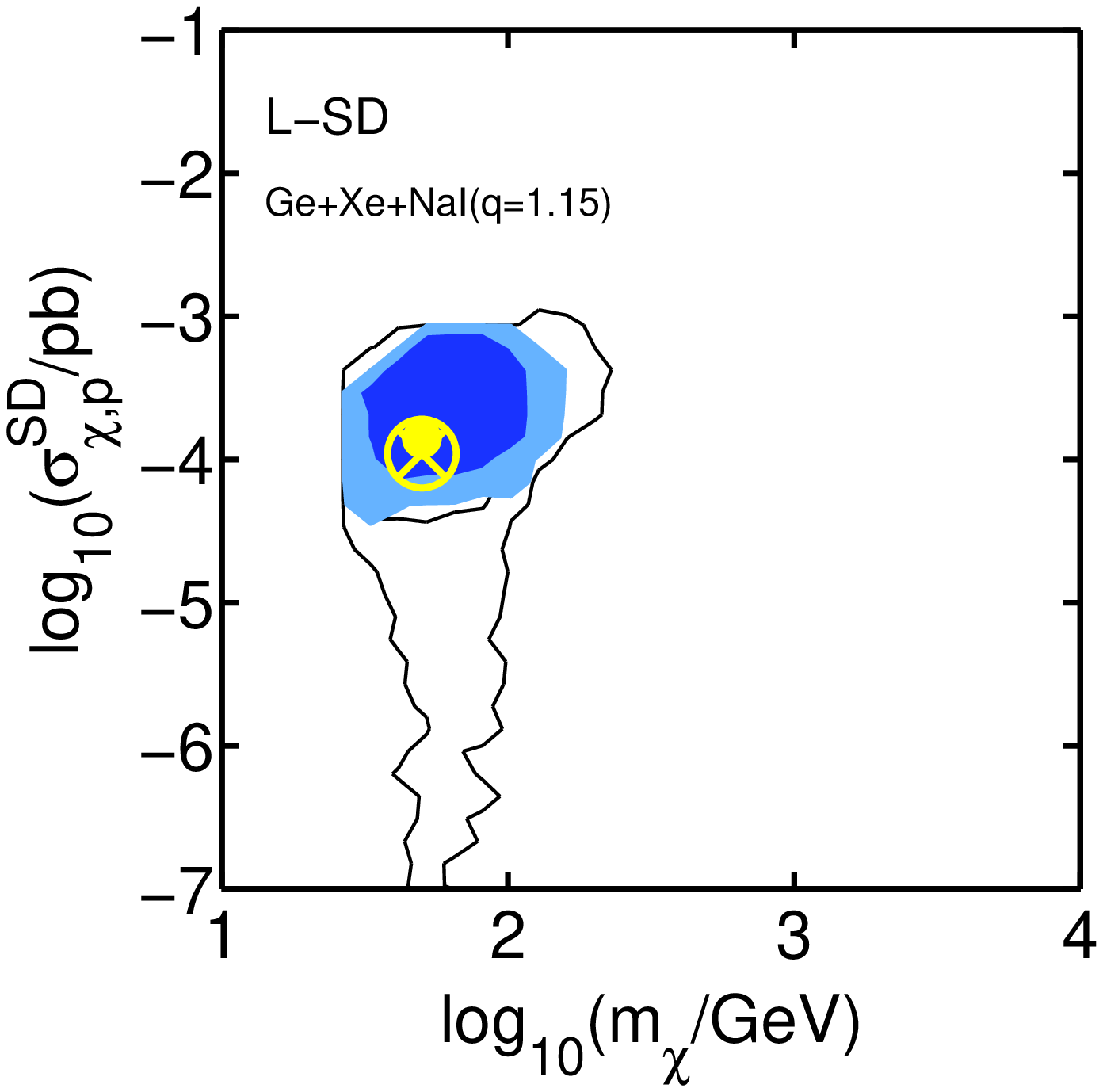}\hspace*{-0.45cm}
    \includegraphics[width=0.37\textwidth]{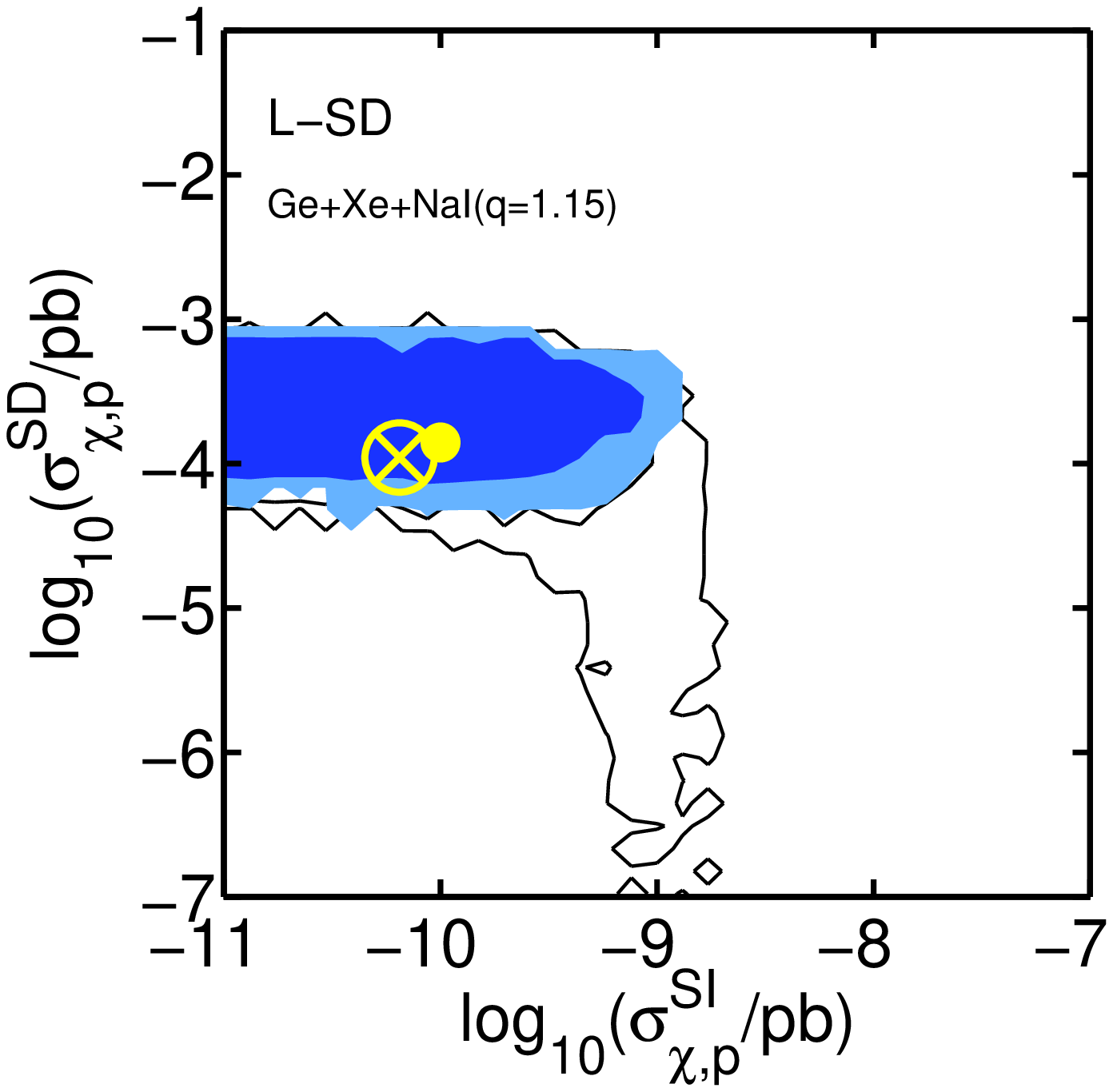}\hspace*{-0.45cm}
\caption{The same as Fig.~\ref{fig:vlsiNaI} but for the L-SD benchmark.
}
\label{fig:lsdNaI}
\end{figure}

Let us now investigate the complementarity potential of scintillating bolometer targets of CaF$_2$ 
and NaI with the Ge and Xe experiments. 
For both bolometric targets, we assume an energy window from 10 to 100 keV, 
a 5\% energy resolution and, as we have done previously for Ge and Xe, a total exposure of 300 kg$\times$yr and a zero background experiment.
Tab.~\ref{tab:bmScint} gives the number of recoil events for each of the bolometric targets. 
In the case of NaI, three different quenching factors have been considered ($q$ = 0.85, 1 and 1.15). 
Following the same procedure of Ref.\,\citen{cerdeno2013a}, for each benchmark and target we have derived the contour plots 
after the combination of data from a Ge detector, a Xe detector and the corresponding bolometric target 
(see Figs.~\ref{fig:vlsiCaF2} to \ref{fig:lsdNaI}). 
Results are shown as blue contours, while black lines correspond to the case when only Ge and Xe are used. 

In benchmark VL-SI, for which Ge and Xe exhibited a degeneracy in the 
($\sigsi$, $\sigsd$) plane, CaF$_2$ provides a good complementarity, allowing the full reconstructions of the 
WIMP parameter space (see Fig.~\ref{fig:vlsiCaF2}). 
This is because for this BM the 97\% (95\%) of the signal in Ge (Xe) is due to the SI component whereas in the case of CaF$_2$, 
a target very sensitive to SD WIMP-nucleon interactions, 
the 80\% of the total rate is due to the SD component. 

\begin{figure}[h]
 \hspace*{-0.5cm}
    \includegraphics[width=0.35\textwidth]{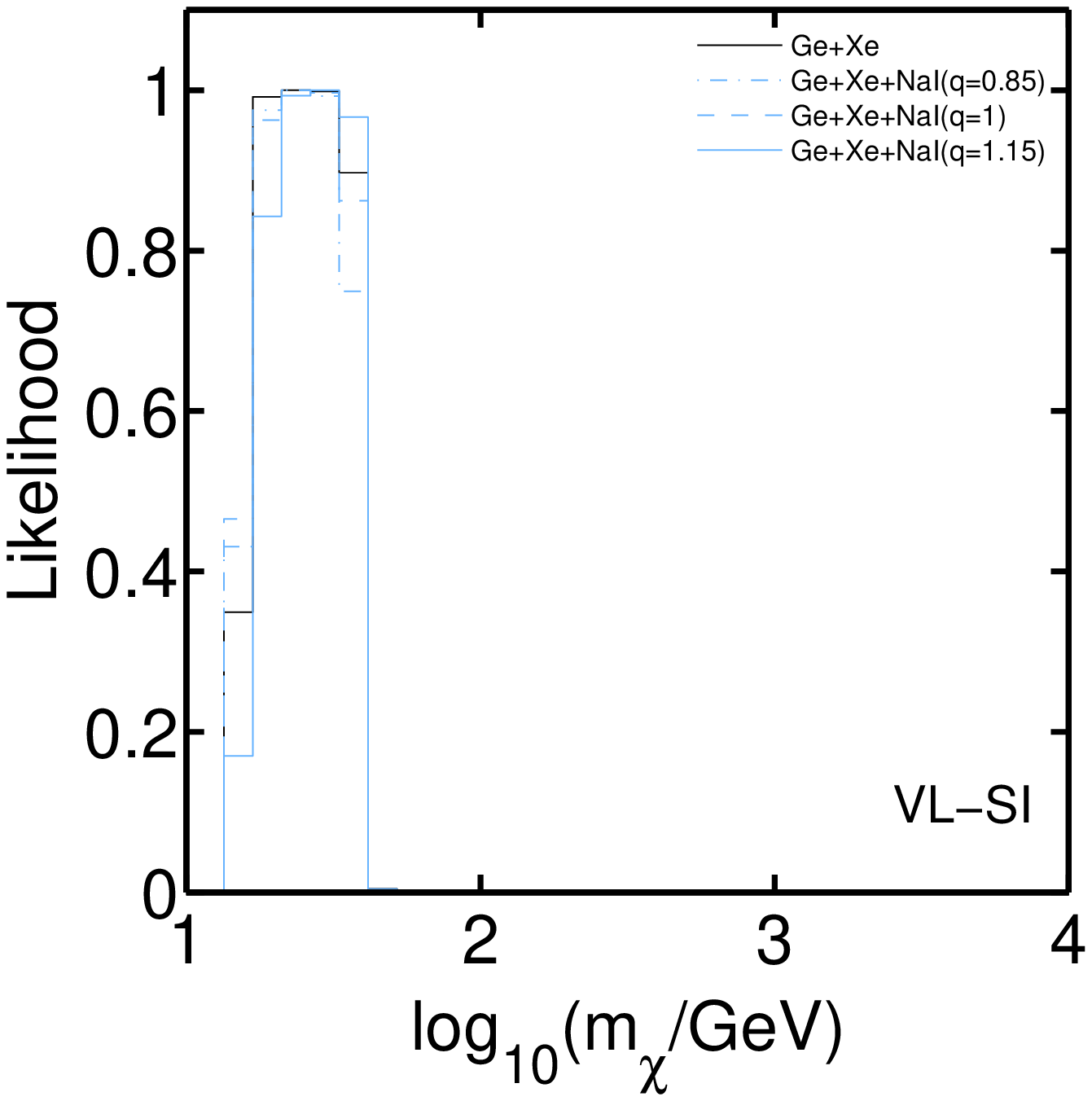}\hspace*{-0.35cm}
    \includegraphics[width=0.35\textwidth]{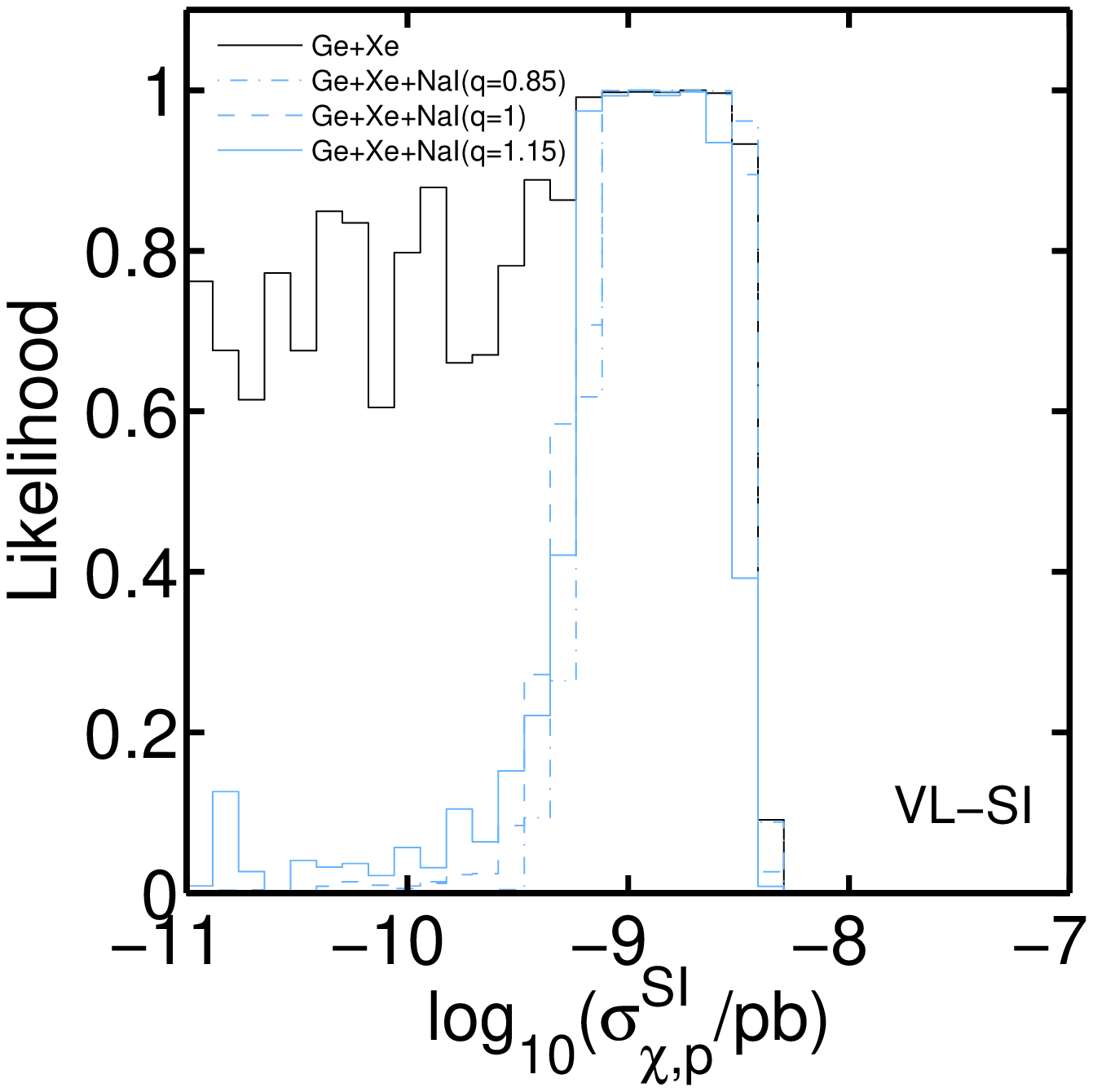}\hspace*{-0.35cm}
    \includegraphics[width=0.35\textwidth]{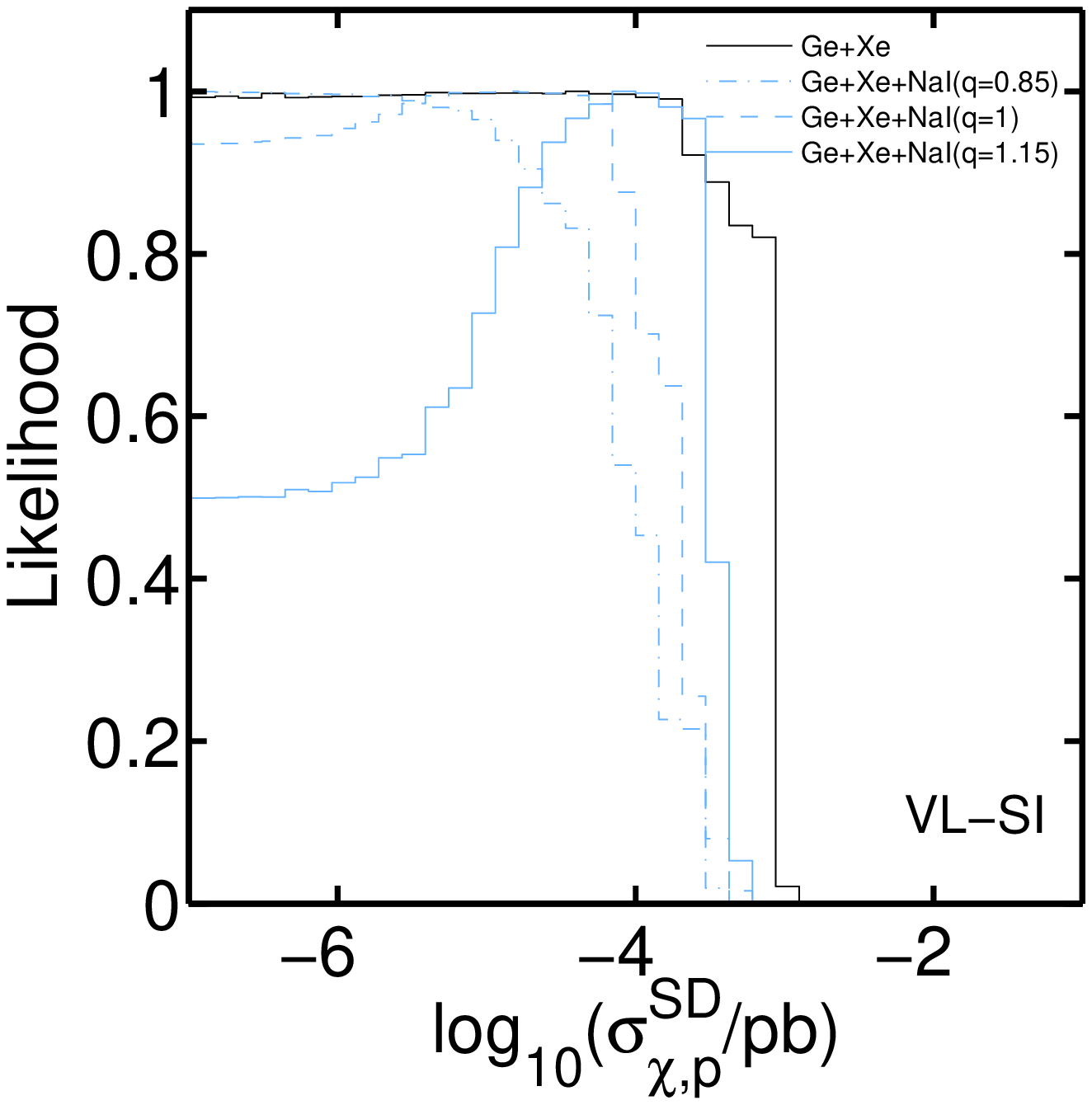}\hspace*{-0.35cm}
\caption{1-D profile likelihood plots for Ge+Xe and Ge+Xe+NaI, considering three different thermal quenching values ($q$=0.85,1,1.15) for 
benchmark VL-SI.
}
\label{fig:vlsi1D}
\end{figure}

The results using NaI are represented in Fig.~\ref{fig:vlsiNaI}, where the three rows correspond to  the three different values considered for the quenching factor.
As we can observe, we are able to obtain closed contours for $\sigsi$, but not for $\sigsd$ (see Fig.~\ref{fig:vlsiNaI}). 
The reason is that, as in the case of Ge and Xe, the signal for this target
is dominated by the SI contribution (approximately 85\% of the total rate).
The change in the quenching factor (which can be understood as a shift in the energy window of nuclear recoils) leads to variations in the number of events due to SD and SI interactions. More importantly these do not change by the same amount, since the energy dependence of the SD and SI form factors is different.
For NaI we observe that the relative contribution due to the SD term increases as the quenching factor decreases, shifting from 14\% at $q$=1.15 to 17\% at $q$=0.85.
This implies that, for this benchmark, the complementarity with Ge and Xe is better for $q=0.85$, as we can observe in Fig.~\ref{fig:vlsiNaI}.
The effect is clearer in the 1-D profile likelihood of the SD cross section shown in Fig.~\ref{fig:vlsi1D}.
Notice also that, although the upper limit on $\sigsd$ is more stringent for $q$=0.85, 
the derived 1-D profile likelihood is practically flat (Fig.~\ref{fig:vlsi1D} right) 
and that leads to a failure in the estimation of $\sigsd$ by the best-fit point (Fig.~\ref{fig:vlsiNaI}).

\begin{figure}[h]
 \hspace*{-0.5cm}
    \includegraphics[width=0.35\textwidth]{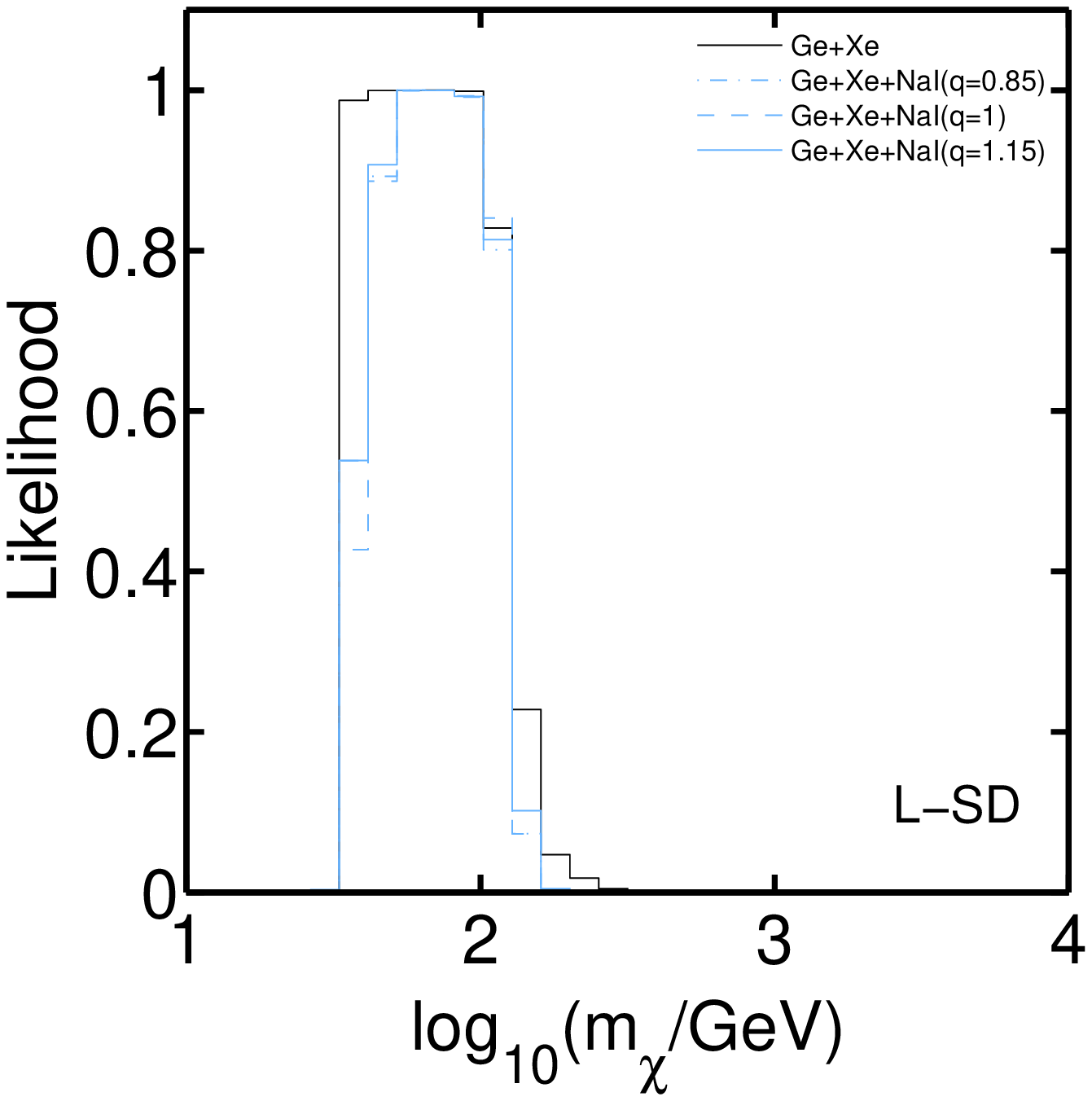}\hspace*{-0.35cm}
    \includegraphics[width=0.35\textwidth]{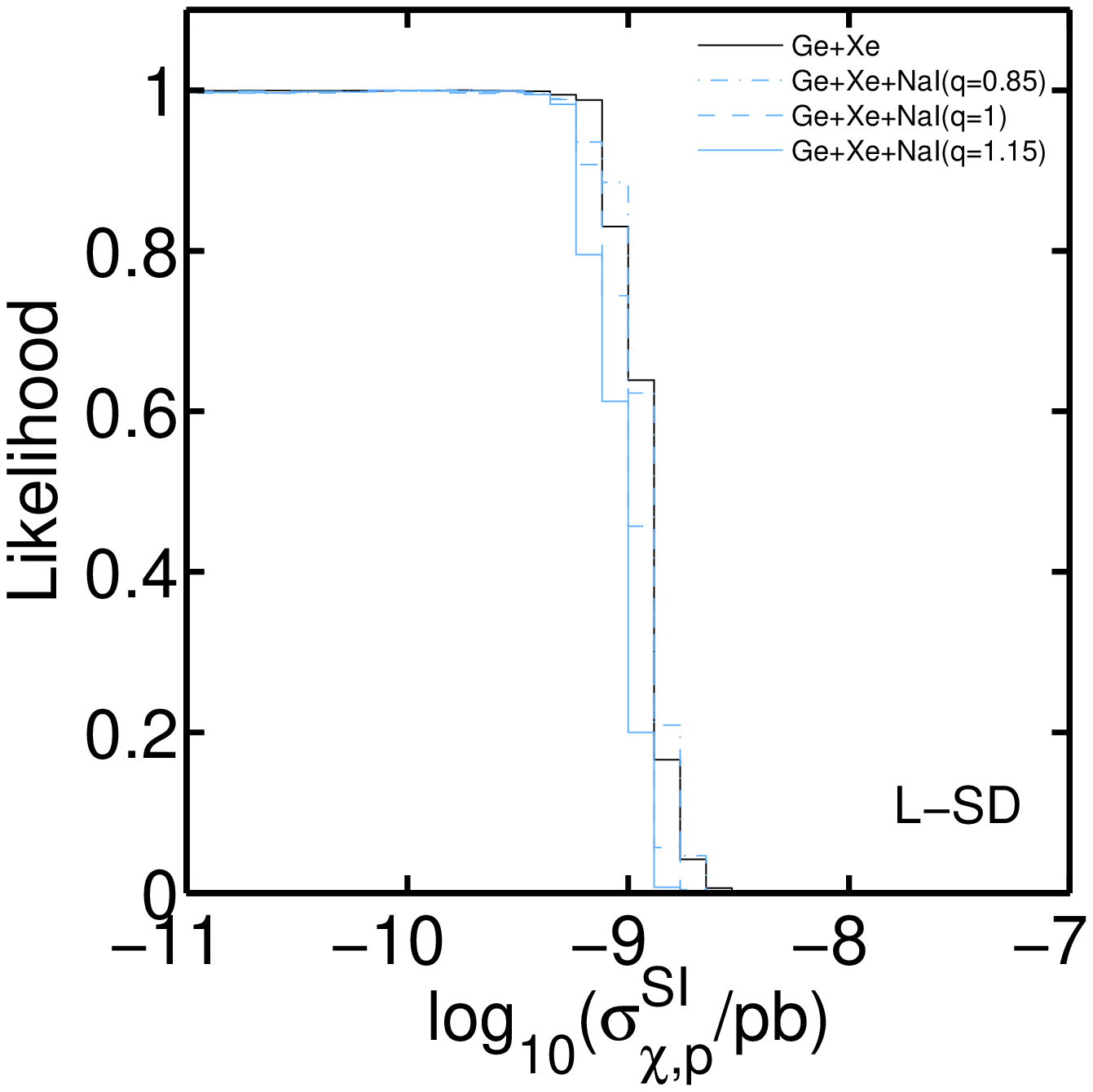}\hspace*{-0.35cm}
    \includegraphics[width=0.35\textwidth]{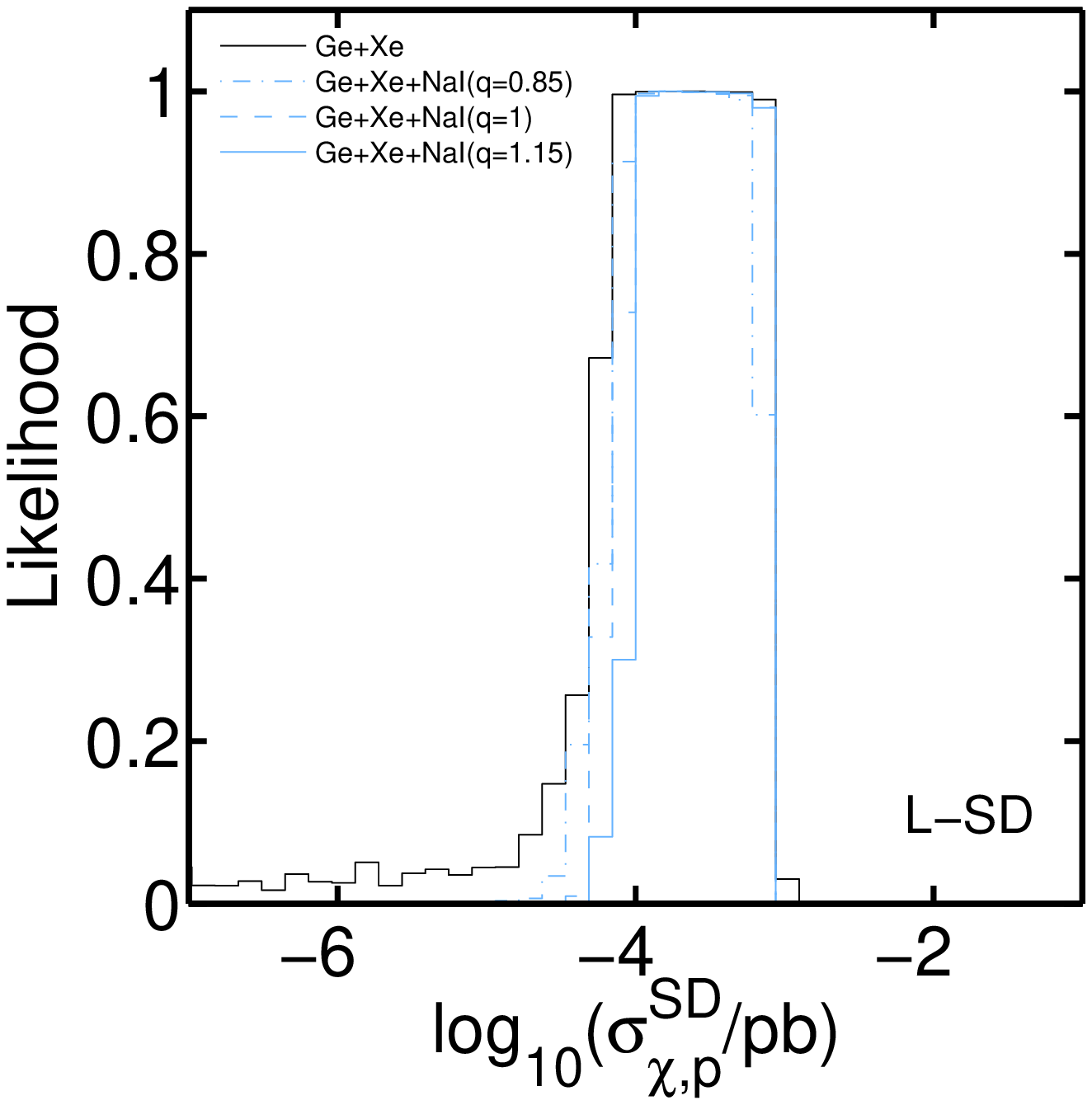}\hspace*{-0.35cm}
\caption{The same as Fig.~\ref{fig:vlsi1D} for the L-SD benchmark.
}
\label{fig:lsd1D}
\end{figure}

The results for benchmark L-SD are shown in  Fig.~\ref{fig:lsdCaF2} for the combination of data from Xe, Ge, and CaF$_2$, and in Fig.~\ref{fig:lsdNaI} for NaI.
As we can see in Tables ~\ref{tab:bm} and ~\ref{tab:bmScint}, in this benchmark the WIMP interactions are dominated by the SD contribution for all the targets. 
Consequently, the degeneracy in the ($\sigsi$, $\sigsd$) plane is not completely removed, although the contours are substantially reduced with respect to the case with Ge and Xe alone. In particular, closed contours appear for $\sigsd$ around the nominal value with both CaF$_2$ and NaI, but only an upper bound for $\sigsi$ is obtained.
In this benchmark the effect of the quenching factor is quite imperceptible 
(see also Fig.~\ref{fig:lsd1D}) because the relative contribution of the SD term is practically 
the same (approximately 7.5\%) for the three values of $q$.

\section{Conclusions}
\label{sec:conclusions}

Following the work done in Ref.\,\citen{cerdeno2013a}, where we investigated the determination of WIMP 
parameters ($\mwimp$, $\sigsi$, $\sigsd$) from a hypothetical direct DM detection with multiple targets, in this paper we have extended the analysis to consider the effect of lower thresholds in Ge and Xe targets, as well as the complementarity potential of two new bolometric targets:  CaF$_2$ and NaI.

We first considered the combination of data from Ge and Xe targets, for both of which we assumed a low energy threshold of $3$~keV to account for recent or projected experimental improvements.
We studied two benchmark scenarios, featuring a very light WIMP ($\mwimp$=20~GeV, $\sigsi$=10$^{-9}$~pb, $\sigsd$=10$^{-5}$~pb)
in which SI contribution dominates the detection rate in both Ge and Xe,
and a light WIMP ($\mwimp$=50~GeV, $\sigsi$=10$^{-10}$~pb, $\sigsd$=1.5$\times$10$^{-4}$~pb)
in which the SD contribution dominates.
Although the combination of data from both targets allows a significant improvement in the reconstruction of DM parameters, a degeneracy in the ($\sigsi$, $\sigsd$) plane usually remains in the points in the parameter space where 
both targets have similar SI/SD ratios.

Scintillating bolometers, with very good energy threshold and resolution and particle discrimination capability, 
provide a wide choice of absorber materials 
that allows to select interesting targets form the point of view of its complementarity
with other experiments. 
In Ref.\,\citen{cerdeno2013a} we studied how certain bolometric targets (CaWO$_4$, Al$_2$O$_3$ 
and LiF) could provide complementary information to data from Ge or Xe based experiments. In this work we have extended the analysis to other two 
scintillating targets (CaF$_2$ and NaI), and considered also the effect of an uncertainty 
in the thermal quenching factor of $\pm$15\%. Both targets are sensitive to the
SD component of the WIMP-nucleus interaction (particularly CaF$_2$ thanks to the presence of $^{19}$F).

We have shown how the inclusion of one of these targets together with 
Ge and Xe can help breaking the degeneracy in the ($\sigsi$, $\sigsd$) plane. 
In particular, in the points of the parameter space for which the rate in Ge and Xe is dominated by the SI contribution and the rate in  CaF$_2$ is mostly SD, the three DM parameters can be reconstructed.
In other examples, although the degeneracy cannot completely removed, at least one of the components of the WIMP-nucleus scattering cross section can be determined. 

We have also shown how a small uncertainty in the thermal quenching factor 
can modify noticeably the parameter reconstruction.

\section*{Acknowledgments}
D.G. Cerde\~no is supported by the Ram\'on y Cajal program of the Spanish MICINN. M. Fornasa is supported by a
Leverhulme Trust grant.  M. Peir\'o is supported by a MultiDark Scholarship. 
Y. Ortigoza is supported by a MultiDark Fellowship. We also 
thank the support of the Consolider-Ingenio 2010 programme under grant MULTIDARK CSD2009-00064, the Spanish MICINN under Grant No. FPA2012-34694, the Spanish MINECO ``Centro de excelencia Severo Ochoa Program" under Grant No. SEV-2012-0249, the Community of Madrid under Grant No. HEPHACOS S2009/ESP-1473, 
the Spanish and the European Regional Development
Fund MINECO-FEDER under grant FPA2011-23749, the Government of Arag\'on,
and the European Union under the Marie Curie-ITN Program No. PITN-GA-2009-237920.
We gratefully acknowledge the access to the High Performance Computing Facility of the University of Nottingham and to the Computing Facilities at the Instituto de F\'isica Te\'orica. 

%
%

\bibliographystyle{ws-ijmpa}
\bibliography{IJMPASpecialIssue}

\end{document}